\documentclass[10pt,final]{IEEEtran}

\usepackage[dvips]{graphicx}
\usepackage{float}
\usepackage{times}
\usepackage{amsmath}
\usepackage{amsfonts}
\usepackage{balance}
\usepackage{color}
\usepackage[normalem]{ulem}
\usepackage{cite}
\renewcommand{\baselinestretch}{1.0}

\renewcommand{\baselinestretch}{1.0}

\begin{document}
%
\title{Bayesian Compressive Sensing Approaches\\ for Direction of Arrival Estimation\\ with Mutual Coupling Effects}



\author{\IEEEauthorblockN{Matthew Hawes$^{a}$, Lyudmila Mihaylova$^{a}$ and Fran\c{c}ois Septier$^{b}$, Simon Godsill$^{c}$}\\[0.3cm]
\IEEEauthorblockA{$^{a}$ Department of Automatic Control and Systems Engineering, University of Sheffield, S1 3JD, UK\\
$^{b}$ IMT Lille Douai, Univ. Lille, CNRS, UMR 9189 - CRIStAL, F-59000 Lille, France \\
$^{c}$ Engineering Department, University of Cambridge, UK\\
{\{m.hawes, l.s.mihaylova\}@sheffield.ac.uk, francois.septier@telecom-lille.fr, sjg@eng.cam.ac.uk }}

}

\markboth{If citing this work please cite the version published in IEEE Transactions on Antennas and Propagation, DOI 10.1109/TAP.2017.2655013}%
{Hawes \MakeLowercase{\textit{et al.}}: Bayesian Compressive Sensing}

\maketitle


\begin{abstract}
The problem of estimating the dynamic direction
of arrival of far field signals impinging on a uniform linear array,
with mutual coupling effects, is addressed. This work proposes two novel
approaches able to provide accurate solutions, including at the endfire
regions of the array. Firstly, a Bayesian compressive sensing
Kalman filter is developed, which accounts for the predicted
estimated signals rather than using the traditional sparse prior.
The posterior probability density function of the received source
signals and the expression for the related marginal likelihood
function are derived theoretically. Next, a Gibbs sampling based
approach with indicator variables in the sparsity prior is
developed. This allows sparsity to be explicitly enforced in
different ways, including when an angle is too far from the previous
estimate. The proposed approaches are validated and evaluated over
different test scenarios and compared to the traditional relevance
vector machine based method. An improved accuracy in terms
of average root mean square error values is achieved (up to 73.39$\%$ for the
modified relevance vector machine based approach and 86.36$\%$ for
the Gibbs sampling based approach).  The proposed approaches prove
to be particularly useful for direction of arrival estimation when
the angle of arrival moves into the endfire region of the array.
\end{abstract}
\section{Introduction}

Direction of arrival (DOA) estimation is the process of determining which direction a signal impinging on an array has arrived from.  Commonly used methods of solving this problem are: MUSIC \cite{Schmidt86,Swindlehurst92}, ESPRIT \cite{Roy89,Zolt96,Tayem04,Gao05} and the maximum likelihood DOA estimator \cite{Zisk88,YDH96,Stocia99}.  However, these methods have some disadvantages, in particular they require knowledge of the number of signals present beforehand and evaluation of a covariance matrix of the array output (adding computational complexity).

Compressive Sensing (CS) theory says that when certain conditions are met it is possible to recover
signals from fewer measurements than used by traditional
methods \cite{Candes06,Donoho06}.  Hence, CS can be applied to the problem of DOA estimation \cite{Mali05,Hyder09,Bilik12,Shen14} by splitting the angular region into $N$ potential DOAs, where only $L<<N$ of the DOAs have an impinging signal (alternatively $N-L$ of the angular directions have a zero-valued impinging signal present).  These DOAs are then estimated by finding the minimum number of DOAs with a non-zero valued impinging signal that still give an acceptable estimate of the array output.

The problem can also be converted into a probabilistic form and solved via Bayesian compressive sensing (BCS) \cite{Ji08}, implemented with a relevance vector machine (RVM)~\cite{Tipping01,Tipping03,Bishop}.  Such a method has been used to solve the problem of static DOA estimation \cite{Carlin13,Yang13}, where a belief of having a sparse received signal is made and the most likely values found.

The Kalman filter (KF) can be used to track dynamic DOAs, with the angular range narrowed to focus in more closely on the DOA estimate from the previous iteration \cite{Khom10}.  However, this prevents directly working with the measured array signals and introduces an additional stage of having to reevaluate the steering vector of the array at each iteration of the KF.  Hierarchical KFs  have been used to track dynamic sparse signals \cite{Karseras13,Filos13}, where the predicted mean of the signals at each iteration is taken as the estimate from the previous iteration and the hyperparameters are estimated using BCS, hence the term Bayesian compressive sensing Kalman Filter (BCSKF).

However, a problem remains when a BCSKF is applied to dynamic DOA estimation with a uniform linear array (ULA).  The estimation accuracy can be reduced when the DOA approaches the endfire region of the array, i.e. when the impinging signal arrives parallel to or almost parallel to the array.  This can be particularly problematic when there is a lot of noise present.

An additional challenge to address when considering the DOA estimation problem is that of mutual coupling.   One way of modeling the mutual coupling effects is to use a mutual coupling matrix~\cite{Su01,Liao12}.  In \cite{Su01} the mutual coupling matrix is found using two methods: minimum mean-square matching and the mutual impedance method.  The method in \cite{Liao12} applies a symmetric Toeplitz matrix, where only antennas within a set separation of each other can cause mutual coupling effects.  In this work the method in \cite{Liao12}  is used to ensure mutual coupling effects are included in the signal model.

The contributions of this paper are: \emph{i}) A BCSKF with a modified RVM, where the traditional sparsity prior is replaced with a belief that the estimated signals will instead match predicted signal values, is proposed.  The result of this new prior is that a new posterior distribution and marginal likelihood have been derived.  Initial results for this method using a signal model without mutual coupling have been reported in  \cite{Hawes15a}. \emph{ii}) A Gibbs sampling approach is proposed. In this approach zero valued signals can be explicitly enforced when there is too large a change in DOA in order to alleviate the estimation accuracy problem for the endfire region of the array. \emph{iii}) A comprehensive performance evaluation is provided, with the proposed methods being compared to a BCSKF using the traditional RVM approach.  Significant improvements in terms of the average root mean square error ($RMSE$) values are observed (up to 73.39$\%$ for the BCSKF with modified RVM and up to 86.36$\%$ for the Gibbs sampling approach).

The remainder of this paper is structured in the following manner:
Section \ref{sec:design} gives details of the proposed estimation methods, including the array model with mutual coupling effects (\ref{sub:AM}), the modified RVM framework for BCS (\ref{sub:RVM}), the BCSKF (\ref{sub:BCSKF}) and the Gibbs sampling implementation (\ref{sub:Gibbs}).  In Section \ref{sec:sim} an evaluation of the effectiveness of the proposed approaches is presented and conclusions are
drawn in Section \ref{sec:con}.

\section{Proposed Estimation Methods}\label{sec:design}
\subsection{Array Model}\label{sub:AM}
\begin{figure}
\begin{center}
   \includegraphics[angle=0,width=0.37\textwidth]{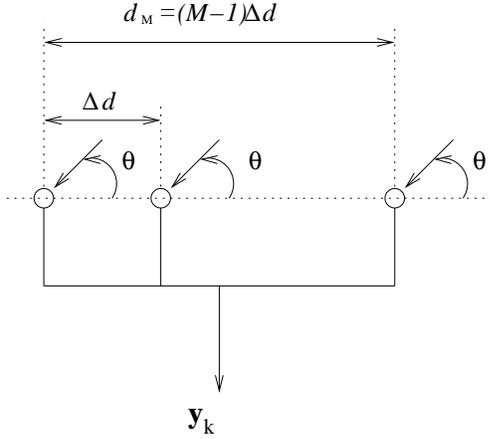}
   \caption{Linear array structure being considered, consisting of $M$ antennas with a uniform adjacent antenna separation of $\Delta d$.
    \label{fig:AM}}
\end{center}
\end{figure}
A narrowband ULA structure consisting of $M$ omnidirectional antennas, with identical responses is shown in
Figure \ref{fig:AM}. Here, a plane-wave signal mode is assumed, i.e. the signal impinges upon the array from the far field and the angle of arrival is limited to $0^{\circ}\leq\theta\leq180^{\circ}$.
The distance from the first antenna to subsequent antennas is
denoted as $d_{m}$ for $m = 1, 2, \ldots, M$, with $d_{1}=0$, i.e.
the distance from the first antenna to itself.  Note, these values are multiples of a uniform adjacent antenna separation of $\Delta d$.

The steering vector of the array is given~by
\begin{equation}
\label{eq:A}
    \textbf{a}(\Omega,\theta)=[1, e^{-j\mu_{2}\Omega
    \cos\theta}, \ldots, e^{-j\mu_{M}\Omega
    \cos\theta}]^{T},
\end{equation}
where $\Omega=\omega T_{s}$ is the normalised frequency with $T_s$
being the sampling period, $\mu_{m}=\frac{d_{m}}{cT_{s}}$ for $m=1,
2, \ldots, M$, $c$ gives the wave propagation speed and $\{\cdot\}^{T}$ denotes the transpose operation.

The array output, $\textbf{y}_{k}$, at time snapshot $k$ is then given by
\begin{equation}\label{eq:y1}
    \textbf{y}_{k} = \textbf{A}_{st}\textbf{x}_{k} + \textbf{n}_{k},
  \end{equation}
where $\textbf{x}_{k}=[x_{k,1}, x_{k,2}, ..., x_{k,N}]^{T}\in\mathbb{C}^{N\times 1}$ gives the received source signals, $\textbf{n}_{k}=[n_{k,1}, n_{k,2}, ..., n_{k,M}]^{T}\in\mathbb{C}^{M\times 1}$ is a noise term, given by a zero mean multivariate Gaussian random variable and $\textbf{A}_{st}=[\textbf{a}(\Omega,\theta_{1}), \textbf{a}(\Omega,\theta_{2}), ..., \textbf{a}(\Omega,\theta_{N})]\in\mathbb{C}^{M\times N}$ is the matrix containing the steering vectors for each angle of interest.  Note, $N$ is the number of points in the grid of potential DOAs the angular region has been split into.  However, only $L<<N$ of these angular directions will have an impinging signal present.

In practice there will also be mutual coupling effects present, which alter the pattern of an individual antenna as compared to if it was being used on its' own.  As a result \eqref{eq:y1} has to be altered to account for this fact. A mutual coupling matrix is used to achieve this \cite{Liao12}, by giving the true steering vector matrix as
\begin{equation}\label{eq:M1}
  \textbf{A}=\textbf{M}_{MC}\textbf{A}_{st}.
\end{equation}
Here $\textbf{M}_{MC}\in\mathbb{C}^{M\times M}$ is the mutual coupling matrix given by
\vspace*{.5cm}
\begin{equation}\label{eq:M2}
  \textbf{M}_{MC} = \left(
                 \begin{array}{cccccc}
                   1 & m_{2} & \ldots & m_{D-1} & \ldots & m_{M} \\
                   m_{2} & 1 & m_{2} & \ldots & \ddots& \vdots \\
                   \vdots & m_{2} & 1 & m_{2} & \ldots & m_{D-1} \\
                   m_{D-1} & \ldots & \ddots & \ddots & \ddots & \vdots \\
                   \vdots & \ddots & \ldots & m_{2} & 1 & m_{2} \\
                   m_{M} & \ldots & m_{D-1} & \ldots & m_{2} & 1 \\
                 \end{array}
               \right).\vspace*{.5cm}
\end{equation}

In \eqref{eq:M2} the mutual coupling coefficients are given by $m_{i}~=~\rho_{i}\exp\{j\phi_{i}\}$ for $i=2, ..., D-1, D, ..., M$, where $\rho_{i}$ and $\phi_{i}$ give the amplitude and phase, respectively.  The variable $D$ places a limit on the separation between antennas above which there will be no mutual coupling effects.  In other words when $i>D$, then $\rho_{i}=0$. This then gives the following:
 \begin{eqnarray}\nonumber
   \textbf{y}_{k} &=& \textbf{M}_{MC}\textbf{A}_{st}\textbf{x}_{k} + \textbf{n}_{k} \\ \label{eq:y}
    &=& \textbf{A}\textbf{x}_{k} + \textbf{n}_{k}.
 \end{eqnarray}

Equation \eqref{eq:y} can then be split into real and imaginary components (given by $\mathcal{R}(\cdot)$ and $\mathcal{I}(\cdot)$, respectively) as follows
\begin{align}\label{eq:y split}
  \nonumber
    \tilde{\textbf{y}}_{k} &= \tilde{\textbf{A}}\tilde{\textbf{x}}_{k}+\tilde{\textbf{n}}_{k} \\
    \left[
      \begin{array}{c}
        \mathcal{R}(\textbf{y}_{k}) \\
        \mathcal{I}(\textbf{y}_{k}) \\
      \end{array}
    \right]
     &= \left[
           \begin{array}{cc}
             \mathcal{R}(\textbf{A}) & -\mathcal{I}(\textbf{A}) \\
             \mathcal{I}(\textbf{A}) & \mathcal{R}(\textbf{A}) \\
           \end{array}
         \right] \left[
                  \begin{array}{c}
                    \mathcal{R}(\textbf{x}_{k}) \\
                    \mathcal{I}(\textbf{x}_{k}) \\
                  \end{array}
                \right] + \left[
                         \begin{array}{c}
                           \mathcal{R}(\textbf{n}_{k}) \\
                           \mathcal{I}(\textbf{n}_{k}) \\
                         \end{array}
                       \right].
  \end{align}
The difference between $\textbf{y}_{k}$ and $\tilde{\textbf{y}}_{k}$ is that $\textbf{y}_{k}$ has been split into its real and imaginary components in $\tilde{\textbf{y}}_{k}$.  As a result the dimensions of $\tilde{\textbf{y}}_{k}$ are increased.  A similar relationship exists between $\textbf{A}$ and $\tilde{\textbf{A}}$, $\textbf{x}_{k}$ and $\tilde{\textbf{x}}_{k}$ and $\textbf{n}_{k}$ and $\tilde{\textbf{n}}_{k}$.

\subsection{Modified Relevance Vector machine for DOA Estimation}\label{sub:RVM}
The aim is to now find a solution for $\tilde{\textbf{x}}_{k}$ which gives the closest possible match to a predicted set of signal values.  To achieve this one can follow a modified RVM framework \cite{Hawes15a}, by evaluating the following
\begin{equation}\label{eq:xest1}
    \tilde{\textbf{x}}_{k,opt} = \max\mathcal{P}(\tilde{\textbf{x}}_{k},\sigma_{k}^{2},\textbf{p}_{k}|\tilde{\textbf{y}}_{k},\tilde{\textbf{x}}_{p}),
  \end{equation}
  where $\sigma_{k}^{2}$ is the variance of the Gaussian noise $\textbf{n}_{k}$, $\textbf{p}_{k}=[p_{k,1}, p_{k,2}, ..., p_{k,2N}]^{T}$ contains the hyperparameters that are to be estimated and $\tilde{\textbf{x}}_{p}=[\mathcal{R}(\textbf{x}_{p})^{T}, \mathcal{I}(\textbf{x}_{p})^{T}]^{T}=[\mathcal{R}(x_{p,1}), ..., \mathcal{R}(x_{p,N}), \mathcal{I}(x_{p,1}), ..., \mathcal{I}(x_{p,N})]^{T}$ holds the predicted values of $\tilde{\textbf{x}}_{k}$.

From (\ref{eq:y split}) it is possible to find:
\begin{equation}\label{eq:y dist}
  \mathcal{P}(\tilde{\textbf{y}}_{k}|\tilde{\textbf{x}}_{k},\sigma_{k}^{2}) = (2\pi\sigma_{k}^{2})^{-M}\exp\Big\{-\frac{1}{2\sigma^{2}}||\tilde{\textbf{y}}_{k} - \tilde{\textbf{A}}\tilde{\textbf{x}}_{k}||_{2}^{2}\Big\}.
\end{equation}
The traditional RVM would now apply a belief that $\tilde{\textbf{x}}_{k}$ is sparse.  However, here this is changed to a belief that $\tilde{\textbf{x}}_{k}$ will match the predicted signals $\tilde{\textbf{x}}_{p}$:
\begin{eqnarray}\label{eq:x dist}\nonumber
  \mathcal{P}(\tilde{\textbf{x}}_{k}|\textbf{p}_{k},\tilde{\textbf{x}}_{p}) &= & (2\pi)^{-N}|\textbf{P}_{k}|^{\frac{1}{2}}\\ &\times&\exp\Big\{-\frac{1}{2}(\tilde{\textbf{x}}_{k} - \tilde{\textbf{x}}_{p})^{T}\textbf{P}_{k}(\tilde{\textbf{x}}_{k} - \tilde{\textbf{x}}_{p})\Big\}.
\end{eqnarray}
Note, when $\tilde{\textbf{x}}_{p} = [0, 0, ..., 0]^{T}$ then (\ref{eq:x dist}) reverts to the hierarchical prior used in the traditional RVM \cite{Ji08,Tipping01} and $|\textbf{P}_{k}|$ indicates the determinant of $\textbf{P}_{k}$, where $\textbf{P}_{k}=\text{diag}(\textbf{p}_{k})$.

It is also necessary to define the hyperparameters over $\textbf{p}_{k}$ and $\sigma_{k}^{2}$.  There are various possibilities for the structuring of the priors on $\textbf{p}_{k}$, which represent mixing parameters in a scale mixture of normals representation of the marginal distribution of $\textbf{x}_{k}$, which will here be in the Student-t family, see e.g. \cite{And74}.  One possibility would be to treat the complex components of $\textbf{x}_{k}$ as complex Student-t distributed, as detailed in \cite{Wolfe02,Wolfe04}. However, this work treats the real and imaginary components of $\textbf{x}_{k}$ as independent Student-t distributed random variables, and hence there are independent Gamma priors for the mixing variables $p_{k,n}$ over all real and imaginary components of $\textbf{x}_{k}$:
 \begin{equation}\label{eq:bcs8}
  \mathcal{P}(\textbf{p}_{k})=\prod_{n=1}^{2N}G(p_{k,n}|\beta_{1},\beta_{2}).
\end{equation}
A Gamma prior can also be used for $\sigma_{k}^{2}$
\begin{equation}\label{eq:bcs9}
  \mathcal{P}(\sigma_{k}^{2})=G(\sigma_{k}^{-2}|\beta_{3},\beta_{4}),
\end{equation}
where $\beta_{1},\beta_{2},\beta_{3}$ and $\beta_{4}$ are scale and shape priors.

It is known that
\begin{equation}\label{eq:BCS1}
  \mathcal{P}(\tilde{\textbf{x}}_{k},\sigma_{k}^{2},\textbf{p}_{k}|\tilde{\textbf{y}}_{k},\tilde{\textbf{x}}_{p})=\mathcal{P}(\tilde{\textbf{x}}_{k}|\tilde{\textbf{y}}_{k},\sigma_{k}^{2},\textbf{p}_{k},\tilde{\textbf{x}}_{p})\mathcal{P}(\sigma_{k}^{2},\textbf{p}_{k}|\tilde{\textbf{y}}_{k},\tilde{\textbf{x}}_{p})
\end{equation}
and
\begin{eqnarray}\label{eq:post}\nonumber
  \mathcal{P}(\tilde{\textbf{x}}_{k}|\tilde{\textbf{y}}_{k},\sigma_{k}^{2},\textbf{p}_{k},\tilde{\textbf{x}}_{p}) = \frac{\mathcal{P}(\tilde{\textbf{y}}_{k}|\tilde{\textbf{x}}_{k},\sigma_{k}^{2})\mathcal{P}(\tilde{\textbf{x}}_{k}|\textbf{p}_{k},\tilde{\textbf{x}}_{p})}{\mathcal{P}(\tilde{\textbf{y}}_{k}|\sigma_{k}^{2},\textbf{p}_{k},\tilde{\textbf{x}}_{p})}\\
  =(2\pi)^{-N}|\boldsymbol\Sigma|^{-1/2}\exp\Bigg\{ -\frac{1}{2}(\tilde{\textbf{x}}_{k}-\boldsymbol\mu)^{T}\boldsymbol\Sigma^{-1}(\tilde{\textbf{x}}_{k}-\boldsymbol\mu)  \Bigg\},
\end{eqnarray}
where the covariance matrix and the mean are given by
\begin{equation}\label{eq:SIGMA}
  \boldsymbol\Sigma = (\sigma_{k}^{-2}\tilde{\textbf{A}}^{T}\tilde{\textbf{A}}+\textbf{P}_{k})^{-1}
\end{equation}
and
\begin{equation}\label{eq:mu}
  \boldsymbol\mu = \boldsymbol\Sigma(\sigma_{k}^{-2}\tilde{\textbf{A}}^{T}\tilde{\textbf{y}}_{k}+\textbf{P}_{k}\tilde{\textbf{x}}_{p}),
\end{equation}
respectively.  Note, the maximum of (\ref{eq:post}) is the posterior
mean~$\boldsymbol\mu$.  For a derivation of \eqref{eq:post} please see Appendix A.

Similarly to \cite{Tipping01}, the probability $\mathcal{P}(\sigma_{k}^{2},\textbf{p}_{k}|\tilde{\textbf{y}}_{k},\tilde{\textbf{x}}_{p})$ can be represented in the following form:
\begin{equation}\label{eq:BCS2}
  \mathcal{P}(\sigma_{k}^{2},\textbf{p}_{k}|\tilde{\textbf{y}}_{k},\tilde{\textbf{x}}_{p})\approx\mathcal{P}(\tilde{\textbf{y}}_{k}|\sigma^{2},\textbf{p},\tilde{\textbf{x}}_{p})\mathcal{P}(\sigma^{2}_{k})\mathcal{P}(\textbf{p}_{k})\mathcal{P}(\tilde{\textbf{x}}_{p}),
\end{equation}
where $\mathcal{P}(\tilde{\textbf{x}}_{p})$ is constant as fixed values are used and the second two terms on the right of are constant if $\beta_{1}=\beta_{2}=\beta_{3}=\beta_{4}=1\times10^{-4}$ as in \cite{Tipping01}.  Therefore, maximising $\mathcal{P}(\sigma_{k}^{2},\textbf{p}_{k}|\tilde{\textbf{y}}_{k},\tilde{\textbf{x}}_{p})$ is roughly equivalent to maximising $\mathcal{P}(\tilde{\textbf{y}}_{k}|\sigma_{k}^{2},\textbf{p}_{k},\tilde{\textbf{x}}_{p})$.  This can be achieved by a type 2 maximisation of its logarithm, which is given by (please see Appendix~B):
\begin{eqnarray}\label{eq:log}\nonumber
  \mathcal{L}(\sigma_{k}^{2},\textbf{p}_{k}) &=& \log\Bigg\{ (2\pi\sigma_{k}^{2})^{-M}|\boldsymbol\Sigma|^{\frac{1}{2}}|\textbf{P}_{k}|^{\frac{1}{2}}\exp\Big(-\frac{1}{2} \\ \nonumber
   &\times&(\tilde{\textbf{y}}_{k}^{T}\textbf{B}\tilde{\textbf{y}}_{k}+\tilde{\textbf{x}}_{p}^{T}\textbf{C}\tilde{\textbf{x}}_{p}-2\sigma_{k}^{2}\tilde{\textbf{y}}_{k}^{T}\tilde{\textbf{A}}\boldsymbol\Sigma\textbf{P}_{k}\tilde{\textbf{x}}_{p})\Big) \Bigg\} \\ \nonumber
   &=& -\frac{1}{2}\Big(2M\log(2\pi)+2M\log\sigma_{k}^{2}-\log|\boldsymbol\Sigma| - \\ \nonumber
   &&\log|\textbf{P}_{k}|+\sigma_{k}^{-2}||\tilde{\textbf{y}}_{k}-\tilde{\textbf{A}}\boldsymbol\mu||^{2}_{2} + \boldsymbol\mu^{T}\textbf{P}_{k}\boldsymbol\mu \\
   &&+\tilde{\textbf{x}}_{p}^{T}\textbf{P}_{k}\tilde{\textbf{x}}_{p}-\tilde{\textbf{x}}_{p}^{T}\textbf{P}_{k}\boldsymbol\mu\Big),
\end{eqnarray}
where $\textbf{B}=(\sigma_{k}^{2}\textbf{I}+\tilde{\textbf{A}}\textbf{P}_{k}^{-1}\tilde{\textbf{A}}^{T})^{-1}$ and $\textbf{C}=\textbf{P}_{k}-\textbf{P}_{k}^{T}\boldsymbol\Sigma\textbf{P}_{k}$.

This is now differentiated with respect to $p_{k,n}$ and $\sigma_{k}^{-2}$ to obtain the update expressions
\begin{equation}\label{eq:pnew}
  p_{k,n}^{new}=\frac{\gamma_{n}}{\mu_{n}^{2}+\tilde{x}_{p,n}^{2}-\tilde{x}_{p,n}\mu_{n}},
\end{equation}
where $\gamma_{n}=1-p_{k,n}\Sigma_{nn}$, $\Sigma_{nn}$ is the $n^{th}$ diagonal element of $\boldsymbol\Sigma$ and
\begin{equation}\label{eq:sigmanew}
  \sigma_{k,new}^{2}=\frac{||\tilde{\textbf{y}}_{k}-\tilde{\textbf{A}}\boldsymbol\mu||_{2}^{2}}{2M-\sum\limits_{n}\gamma_{n}}.
\end{equation}
For the derivation of \eqref{eq:pnew} and \eqref{eq:sigmanew} please see Appendix C.

The maximisation is then achieved by iteratively finding $\boldsymbol\Sigma$ and $\boldsymbol\mu$, followed by $p_{k,n}^{new}$ for $n=1, ..., N$ and $\sigma_{k,new}^{2}$ until a convergence criterion is met \cite{Ji08,Tipping01}.  In other words, the new estimates for the noise variance and precision hyperparameters found from \eqref{eq:sigmanew} and \eqref{eq:pnew} are then used in \eqref{eq:SIGMA} and \eqref{eq:mu} to find new estimates of the covariance matrix and mean of the distribution in \eqref{eq:post}.  Note that when $\tilde{\textbf{x}}_{p} = [0, 0, ..., 0]^{T}$ the update expressions match those used by the traditional RVM.

The final estimate of the received signals is then given by
\begin{equation}\label{eq:xopt}
  \tilde{\textbf{x}}_{k,opt}=\Big(\frac{\tilde{\textbf{A}}^{T}\tilde{\textbf{A}}}{\sigma_{k,opt}^{2}}+\textbf{P}_{k,opt}\Big)^{-1}\Big( \frac{\tilde{\textbf{A}}^{T}\tilde{\textbf{y}}_{k}}{\sigma_{k,opt}^{2}}+\textbf{P}_{k,opt}\tilde{\textbf{x}}_{p} \Big)
\end{equation}
where $\sigma^{2}_{k,opt}$ and $\textbf{P}_{k,opt}=\text{diag}(p_{k,opt,1}, p_{k,opt,2}, ..., p_{k,opt,2N})$ are the result of optimising the noise estimate and hyperparameters, respectively.  Now $\tilde{\textbf{x}}_{k,opt}$ can be used to reconstruct the estimated signals as
\begin{equation}\label{eq:xfinal}
  x_{k,opt,n}=\tilde{x}_{k,opt,n}+j\tilde{x}_{k,opt,N+n},
\end{equation}
where $n=1, 2, ..., N$.

The thresholding scheme in \cite{Carlin13} can then be applied to keep the $\tilde{L}$ most significant signals.  To do this find the total energy content of the estimated received signals and then sort them.  A threshold value, $\eta$, is then defined as a percentage of the energy content that is to be retained.  Starting with the most significant estimated signal, the estimated signals are summed until the threshold is reached and the remaining signals are then set to be equal to 0.  The remaining non-zero valued signals then give the DOA estimates and $\tilde{L}$ is an estimate of the number of far field signals impinging on the array.

\subsection{Bayesian Compressive Sensing Kalman Filter}\label{sub:BCSKF}
In order to track the changes in the DOA estimates at each time snapshot the modified RVM based DOA estimation procedure detailed above is combined with a Bayesian KF, giving a BCSKF for DOA estimation \cite{Hawes15a}.  The signal model described above is again used along with the prediction
\begin{align} \nonumber
    \tilde{\textbf{x}}_{k|k-1} &= \tilde{\textbf{x}}_{k-1|k-1}+\boldsymbol\Delta\textbf{x} & \boldsymbol\Sigma_{k|k-1} &= \boldsymbol\Sigma_{k-1} + \textbf{P}^{-1}_{k} \\
    \tilde{\textbf{y}}_{k|k-1} &= \tilde{\textbf{A}}\tilde{\textbf{x}}_{k|k-1}& \tilde{\textbf{y}}_{e,k} &= \tilde{\textbf{y}}_{k}-\tilde{\textbf{y}}_{k|k-1}
  \end{align}
and update steps
  \begin{align} \nonumber
  \tilde{\textbf{x}}_{k} = \tilde{\textbf{x}}_{k|k-1} + \textbf{K}_{k}\tilde{\textbf{y}}_{e,k}&\;\;\; \boldsymbol\Sigma_{k|k}=(\textbf{I}-\textbf{K}_{k}\tilde{\textbf{A}})\boldsymbol\Sigma_{k|k-1}\\
   \textbf{K}_{k}=\boldsymbol\Sigma_{k|k-1}\tilde{\textbf{A}}^{T}&(\sigma_{k}^{2}\textbf{I}+\tilde{\textbf{A}}\boldsymbol\Sigma_{k|k-1}\tilde{\textbf{A}}^{T})^{-1}
  \end{align}
of the BCSKF.  Here, $k|k-1$ indicates prediction at time instance $k$ given the previous measurements and $\boldsymbol\Delta\textbf{x}$ is determined by the assumed DOA change.  Note, $\boldsymbol\Delta\textbf{x}$ is fixed by the predetermined constant motion rather than being a random noise term. For example, if the angular range is sampled every $1^{\circ}$ and the DOA is assumed to increase by $2^{\circ}$ then $\boldsymbol\Delta\textbf{x}$ will be selected to increase the index of the non-zero valued entries in $\tilde{\textbf{x}}_{k-1|k-1}$ by two to give the index of the non-zero valued entries in $\tilde{\textbf{x}}_{k|k-1}$.

At each time snapshot it is necessary to estimate the noise variance and hyperparameters in order to evaluate the prediction and update steps of the BCSKF.  This is done by considering the log likelihood function given by
\begin{eqnarray}\nonumber
  \mathcal{L}(\sigma_{k}^{2},\textbf{p}_{k}) &=& -\frac{1}{2}\Big(2M\log(2\pi)+2M\log\sigma_{k}^{2}-\log|\boldsymbol\Sigma|\\ \nonumber&& -
   \log|\textbf{P}_{k}|+\sigma_{k}^{-2}||\tilde{\textbf{y}}_{e,k}-\tilde{\textbf{A}}\boldsymbol\mu||^{2}_{2} + \boldsymbol\mu^{T}\textbf{P}_{k}\boldsymbol\mu \\
   &&+\tilde{\textbf{x}}_{k|k-1}^{T}\textbf{P}_{k}\tilde{\textbf{x}}_{k|k-1}-\tilde{\textbf{x}}_{k|k-1}^{T}\textbf{P}\boldsymbol\mu\Big),
\end{eqnarray}
which can be optimised by following the procedure described in Section \ref{sub:RVM}.  In other words we apply the modified RVM framework to $\tilde{\textbf{y}}_{e,k}$, using the KF prediction $\tilde{\textbf{x}}_{k|k-1}$ as the expected estimate values $\tilde{\textbf{x}}_{p}$.

It is worth noting that the continued accuracy of the proposed BCSKF relies on the accuracy of the initial estimate and the parameter values selected.  If the initial estimate (made using the framework described in Section \ref{sub:RVM} and $\tilde{\textbf{x}}_{p}=[0, 0, ..., 0]^{T}$) of the received signals is accurate and sparse, then the priors that are enforced will ensure this continues to be the case.  However, an inaccurate initial DOA estimate or poorly matched expected DOA change can lead to the introduction of inaccuracies in subsequent estimates.  Similarly, if the initial estimate of the received signals is not sparse then subsequent estimates are likely to not be sparse.  As a result, care should be taken when choosing the initial parameter values and determining the likely DOA change.

\subsection{Gibbs Sampling for DOA Estimation}\label{sub:Gibbs}
The method described in the previous sections based on a BCSKF with a modified RVM required the use of prior knowledge of the predicted change in DOA.  However, in practice this may not always be known, making it important to have an alternative method that can still give improved accuracy for the endfire region.

This work proposes using a sparsity prior which is given as a combination of a point mass concentrated at zero (Dirac delta function) and a zero mean Gaussian distribution, \cite{Fev06,He09,Yu12}, giving
\begin{equation}\label{eq:xpz}
  \mathcal{P}(\tilde{\textbf{x}}_{k}|\textbf{p}_{k},\tilde{\textbf{z}}_{k}) = \prod\limits_{n=1}^{2N}(1-\tilde{z}_{k,n})\delta_{0}+\tilde{z}_{k,n}\mathcal{N}(\tilde{x}_{k,n}|0,p_{k,n}),
\end{equation}
where $\tilde{\textbf{z}}_{k}=[\textbf{z}_{k}^{T}, \textbf{z}_{k}^{T}]^{T}$ and $\textbf{z}_{k}=[z_{k,1}, z_{k,2}, ..., z_{k,N}]^{T}$.

Note, $\tilde{z}_{k,n}$ is the indicator variable for $\tilde{x}_{k,n}$  and determines which of the two components in \eqref{eq:xpz} is selected.  When $\tilde{z}_{k,n}~=~0$, the value of $\tilde{x}_{k,n}$ is determined solely by the point mass concentrated at zero.  As a result, $\tilde{x}_{k,n}=0$ and sparsity is explicitly introduced. Alternatively, when $\tilde{z}_{k,n}=1$ the value of $\tilde{x}_{k,n}$ is determined by the Gaussian distribution allowing a non-zero valued estimate.  The repetition of $\textbf{z}_{k}$ in $\tilde{\textbf{z}}_{k}$ means that the same indicator variable is used for both the real and imaginary parts of each entry in $\textbf{x}_{k}$.

This indicator value can also be used to address the endfire accuracy problem by selecting the value of $z_{k,n}=0$ if $|n-i|>j$.  Here $i$ is the index of the closet non-zero valued estimate from the previous time snapshot and $j$ defines a maximum allowed change in the DOA estimate.  Only $n=1, 2, ..., N$ is considered to get the entries for $\textbf{z}_{k}$, with $\tilde{\textbf{z}}_{k}$ then being found as previously stated.

This leaves  the following
\begin{eqnarray}\label{eq:zn}
 z_{k,n} =
 \begin{cases}
 z_{k,n}^{1} & \text{ if }|n-i|\leq j, \\
 z_{k,n}^{2} & \text{ if }|n-i|>j, \\
 \end{cases}
  \end{eqnarray}
where $z_{k,n}^{1}$ and $z_{n}^{2}$ are defined by the following Beta distributions
\begin{eqnarray}\label{eq:znb1} \nonumber
  z_{k,n}^{1} &=& B(z_{k,n}^{1}|\beta_{5}^{1},\beta_{6}^{1}), \\
  z_{k,n}^{2} &=& B(z_{k,n}^{2}|\beta_{5}^{2},\beta_{6}^{2}).
\end{eqnarray}
In order to enforce zero-valued estimates when $|n-i|>j$, it is necessary to select $\beta_{5}^{2}$ and $\beta_{6}^{2}$ to ensure a zero-valued $z_{k,n}$ is preferred.  However, when $|n-i|\leq j$ it is necessary to choose $\beta_{5}^{1}$ and $\beta_{6}^{1}$ so that the chances of $z_{k,n}=0$ and $z_{k,n}=1$ are equal.  This can be achieved by
\begin{eqnarray}\label{eq:znb2}\nonumber
  \mathcal{P}(z_{k,n}^{1}|\beta_{5},\beta_{6}) &=& B(z_{k,n}^{1}|\beta_{5},\beta_{6}), \\
  \mathcal{P}(z_{k,n}^{2}|\beta_{5},\beta_{6}) &=& B\bigg(z_{k,n}^{2}|\beta_{5}-\frac{1}{j},\beta_{6}+\frac{1}{j}\bigg),
\end{eqnarray}
where $\beta_{5}=\beta_{6}=1$.

The posterior distribution of $\tilde{\textbf{x}}_{k}$ can be written as \cite{Yu12}
\begin{eqnarray}\label{eq:postx}\nonumber
  \mathcal{P}(\tilde{\textbf{x}}_{k}|\tilde{\textbf{y}}_{k},\sigma_{k}^{2},\textbf{p}_{k},\tilde{\textbf{z}}_{k})\propto\Big\{\prod\limits_{n=1}^{2N}[(1-\tilde{z}_{k,n})\delta_{0}+\tilde{z}_{k,n}\\
  \times  \mathcal{N}(\tilde{x}_{k,n}|0,p_{k,n})]\Big\}\mathcal{N}(\tilde{\textbf{y}}_{k}|\tilde{\textbf{A}}\tilde{\textbf{x}}_{k},\sigma_{k}^{-2}).
\end{eqnarray}
Now also define $\tilde{\textbf{A}}_{n}$ as being the entries in
$\tilde{\textbf{A}}$ relating to the index $n$ and
$\tilde{\textbf{A}}_{-n}$ are the entries of $\tilde{\textbf{A}}$
excluding the entries relating to index $n$ (and similarly for $\tilde{\textbf{x}}_{k}$).  Then as per Appendix D this gives
\begin{eqnarray}\label{eq:gibbs1}\nonumber
  \mathcal{P}(\tilde{x}_{k,n}|\tilde{\textbf{y}}_{k},\tilde{\textbf{x}}_{k,-n},\sigma_{k}^{2},\textbf{p}_{k},\tilde{\textbf{z}}_{k}) &=& (1-\hat{z}_{k,n})\delta_{0}
  +\hat{z}_{k,n}\\ &\times&\mathcal{N}(\tilde{x}_{k,n}|\hat{\mu}_{k,n},\hat{p}_{k,n}),\\\label{eq:gibbs2}
  \hat{p}_{k,n} &=& p_{k,n}+p_{0}\tilde{\textbf{A}}_{n}^{T}\tilde{\textbf{A}}_{n}, \\\label{eq:gibbs3}
  \hat{\mu}_{k,n} &=& \hat{p}_{k,n}^{-1}p_{0}\tilde{\textbf{A}}_{n}^{T}\tilde{\textbf{y}}_{k,n}, \\ \label{eq:gibbs4} \nonumber
  \frac{\hat{z}_{k,n}}{1-\hat{z}_{k,n}} &=& \frac{\tilde{z}_{k,n}}{1-\tilde{z}_{k,n}}\\
  &\times&\frac{\mathcal{N}(0|0,p_{k,n})}{\mathcal{N}(0|\tilde{\mu}_{k,n},\hat{p}_{k,n})},
\end{eqnarray}
where $p_{0}=1/\sigma_{k}^{2}$ and $\tilde{\textbf{y}}_{k,n}=\tilde{\textbf{y}}_{k}-\tilde{\textbf{A}}_{-n}\tilde{\textbf{x}}_{k,-n}$.

There are two further posterior distributions that have to be considered.  That is the distributions for $p_{k,n}$ and $p_{0}$ which are given by
\begin{equation}\label{eq:pnpost}
  \mathcal{P}(p_{k,n}|\textbf{x}_{k-1}) = G(\beta_{1}+||\textbf{x}_{k-1,nj}||_{0},\beta_{2}+||\textbf{x}_{k-1,nj}||_{2}^{2})
\end{equation}
and
\begin{equation}\label{eq:p0post}
  \mathcal{P}(p_{0}|\tilde{\textbf{y}}_{k},\tilde{\textbf{x}}_{k})=G(\beta_{3}+M,\beta_{4}+\frac{1}{2}||\tilde{\textbf{y}}_{k}-\tilde{\textbf{A}}\tilde{\textbf{x}}_{k}||_{2}^{2}),
\end{equation}
respectively.  Note, in \eqref{eq:pnpost} $\textbf{x}_{k-1,nj}$ gives the entries within $\textbf{x}_{k-1}$ that have an index within the distance $j$ of index $n$.  By using $\textbf{x}$ rather than  $\tilde{\textbf{x}}$ to find $\textbf{x}_{k-1,nj}$ it guarantees the same value of $||\textbf{x}_{k-1,nj}||_{2}^{2}$ and $||\textbf{x}_{k-1,nj}||_{0}$ for both the real and imaginary components.

As a result the Gibbs sampling steps are as detailed below:
\begin{enumerate}
  \item Sample $\tilde{x}_{k,n}$ from $\mathcal{P}(\tilde{x}_{k,n}|\tilde{\textbf{y}}_{k},\tilde{\textbf{x}}_{k,-n},\sigma_{k}^{2},\textbf{p}_{k},\tilde{\textbf{z}}_{k})$.
  \item Sample $p_{k,n}$ from $\mathcal{P}(p_{k,n}|\textbf{x}_{k-1})$.
  \item $\textbf{if }n\leq N$ then Sample $z_{k,n}^{1}$ from $\mathcal{P}(z_{k,n}^{1}|\beta_{5},\beta_{6})$, $\textbf{else } z_{k,n}^{1}=z_{k,n-N}^{1}$.
  \item $\textbf{if }n\leq N$ then Sample $z_{k,n}^{2}$ from $\mathcal{P}(z_{k,n}^{2}|\beta_{5},\beta_{6})$, $\textbf{else } z_{k,n}^{2}=z_{k,n-N}^{2}$.
  \item Sample $p_{0}$ from $\mathcal{P}(p_{0}|\tilde{\textbf{y}}_{k},\tilde{\textbf{x}}_{k})$.
\end{enumerate}

These steps are done for each of the $T$ iterations of the Gibbs sampler, where the first $T_{BI}$ iterations are the burn-in iterations.  The final estimate of the received array signals is then given by the mean values of the final $T-T_{BI}$ iterations \cite{He09}.  The DOA estimate can then be found using the previously described thresholding scheme (see \ref{sub:RVM}), with the remaining non-zero valued estimates corresponding to the DOA estimates.

Note, the performance of this method will again heavily depend on the accuracy of the first estimate.  As a result, it is possible to use the traditional BCS DOA estimation method (Section \ref{sub:RVM} with $\tilde{\textbf{x}}_{p}=[0, 0, ..., 0]^{T}$) to ensure an as accurate as possible intial estimate at the first time snapshot.  The proposed Gibbs sampling based method can then be used at the subsequent time snapshots to get the next DOA estimate.
\section{Performance Evaluation}\label{sec:sim}

In this section a comparison in performance of the proposed methods and the traditional RVM based BCSKF method will be made over five example scenarios, under the same test conditions.   Firstly, an example is considered where the initial DOA starts outside of the endfire region and then moves into it.  Secondly, an example is given where the DOA remains out of the endfire region.  In the third scenario the initial DOAs and the signal values are randomly generated.  Then the evaluation will also consider the scenario where there is a mismatch between the actual and assumed change in DOA.  Finally, the evaluation will consider a random change in DOA at each time snapshot.  This means that $\Delta\textbf{x}$ which is selected for the modified RVM based BCSKF will not be a true reflection of how the DOA actually changes for the last two examples.

Note, the term traditional RVM based BCSKF method means the entries of $\textbf{P}_{k}$ in the prediction step of the BCSKF are found using the RVM optimisation method as detailed in \cite{Ji08,Tipping01}.  In other words this is the method detailed in Section \ref{sub:RVM} with $\tilde{\textbf{x}}_{p}=\boldsymbol0$.  All of the examples are implemented in Matlab on a computer with
an Intel Xeon CPU E3-1271 (3.60GHz) and 16GB of RAM.

The performance of each method will be measured using the $RMSE$ in the DOA estimate.  This is given by
\begin{equation}\label{eq:RMSE}
  RMSE = \sqrt{\frac{\sum\limits_{q=1}^{Q}\sum\limits_{l=1}^{\tilde{L}}|\theta_{l}-\hat{\theta}_{l}|^{2}}{Q\tilde{L}}},
\end{equation}
where $\theta_{l}$ is the actual DOA, $\hat{\theta}_{l}$ is the estimated DOA and $Q$ is the number of Monte Carlo simulations carried out, with $Q=100$ being used in each case.  This gives a measure of the estimation accuracy and the computation time will be used as a measure of the complexity of each method.

For the Gibbs sampling method a burn-in period of 250 iterations is used and then 50 further iterations used to find the final estimate of the received array signals.  When a distance of $j=5^{\circ}$ is exceeded a zero-valued estimate of the received signals is enforced in order to alleviate the endfire accuracy problem.

For all the design examples considered the selection of
 $\sigma_{k}^{2}~=~0.4$  as the noise variance is used,  with an initial estimate
of the noise variance given by $\sigma_{k,0}^{2}=0.1$.  The array
geometry being used is that of a ULA with $M=20$ antennas and an
adjacent antenna separation of $\frac{\lambda}{2}$, where $\lambda$
is the wavelength of the signal of interest.  This gives an array aperture of $9.5\lambda$.  For the mutual coupling matrix a value of $D=3$ is selected, meaning that a separation of $1.5\lambda$ or greater gives negligible mutual coupling effects.  The values $\rho_{1}=0.65$, $\rho_{2}=0.25$, $\phi_{1}=\pi/7$ and $\phi_{2}=\pi/10$ are then also used.  Finally, in each example a single narrowband signal impinging on the array is considered, meaning $L=1$.

\subsection{Endfire Region}
\begin{figure}
\begin{center}
   \includegraphics[angle=0,width=0.45\textwidth]{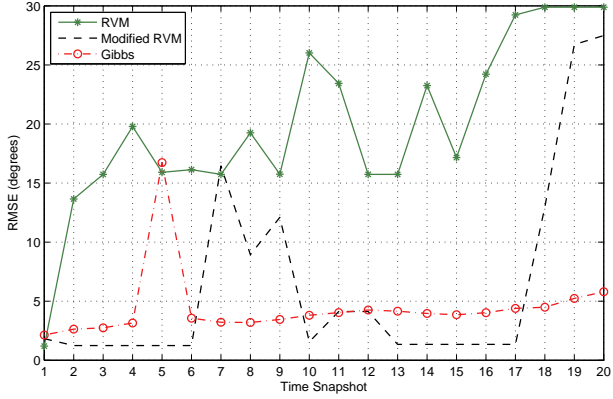}
   \caption{$RMSE$ values for the endfire region example.
    \label{fig:end}}
\end{center}
\end{figure}

For this example the initial DOA of the signal is $\theta=20^{\circ}$, which then decreases by $1^{\circ}$ at each time snapshot.  The signal value at each snapshot is set to be 1.  Table \ref{tb:end} summarises the performance of the three methods for this example, with the $RMSE$ values at each time snapshot being shown in Figure~\ref{fig:end}.

\begin{table}
\caption{Performance summary for the endfire region example.} \centering
\begin{tabular}{|c|c|c|}
  \hline
   & Average $RMSE$  & Average Computation  \\
  Method&(degrees) &Time (seconds) \\
  \hline
  RVM &19.88   & 0.76  \\
  \hline
  Modified RVM & 6.46  & 0.98  \\
  \hline
  Gibbs &  3.82 &  17.83 \text{excluding burn-in}\\
  \cline{3-3}
  & &   107.84 \text{including burn-in} \\
  \hline
\end{tabular}
\label{tb:end}
\end{table}

 Here it can be seen that there has been a significant decrease in the average $RMSE$ values for both the modified RVM method (67.51$\%$ improvement) and the Gibbs sampling based method (80.78$\%$ improvement).  Overall this suggests that a more accurate estimate of the DOA is possible.  It is worth noting that there has still been an increase in the RMSE for the modified RVM based approach in the endfire region of the angular range.  However, this has come much later on the than for the traditional RVM based approach (indicating a degradation in performance for a smaller angular range) and the maximum $RMSE$ value reached is lower (indicating the degradation is less severe).

These improvements have come at the cost of an increased computation time in both instances.  For the modified RVM method this increase is insignificant as the average computation time is still less than one second.  The increase for the Gibbs sampling based method is larger, illustrating an increase in computational complexity.  However, it is worth remembering that this increase has resulted in a more accurate DOA estimate being achieved without prior knowledge about what the change in DOA will be.

In this instance the results suggest that one of the two proposed methods should be used when the estimated DOA approaches the endfire region of the array.  If the change in DOA is known in advance and computational complexity is a primary concern then the modified RVM based method is the most suitable (a more accurate estimate can be achieved without a large increase in computation time).  However, when this information is not available, or computational complexity is not a concern, it is possible to get a significant improvement in accuracy (at the cost of computation time) using the Gibbs sampling based method.

In the previous simulation  an adjacent antenna separation of $\lambda/2$ is used as it is known that this is the largest separation that can be used while still avoiding a degraded performance due to the introduction of grating globes \cite{vantrees02a}.  However, an example of what the relative performance of the methods is when a smaller adjacent antenna separation will now be considered.

In this instance an adjacent antenna separation of $\lambda/4$ is selected.  As the array aperture is kept constant (to allow a fair comparison between adjacent antenna separation sizes) this means the number of antennas is given by $M=39$.  This also means a value of $D=9$ is required to keep the same distance limits on mutual coupling occurring.  The values of $\rho_{i}$ and $\phi_{i}$ are then selected to be uniformly spread over the range of 0.65 to 0.25 and $\pi/7$ to $\pi/10$, respectively.  The remaining parameters are kept constant and the same test scenario as for the previous example is used.

\begin{table}
\caption{Performance summary for the endfire region example with reduced adjacent antenna separations.} \centering
\begin{tabular}{|c|c|c|}
  \hline
   & Average $RMSE$  & Average Computation  \\
  Method&(degrees) &Time (seconds) \\
  \hline
  RVM & 2.21  & 0.98  \\
  \hline
  Modified RVM &  0.86 & 1.25  \\
  \hline
  Gibbs & 3.02  & 16.81  \text{excluding burn-in}\\
  \cline{3-3}
  & &   101.65 \text{including burn-in} \\
  \hline
\end{tabular}
\label{tb:end2}
\end{table}

\begin{figure}
\begin{center}
   \includegraphics[angle=0,width=0.45\textwidth]{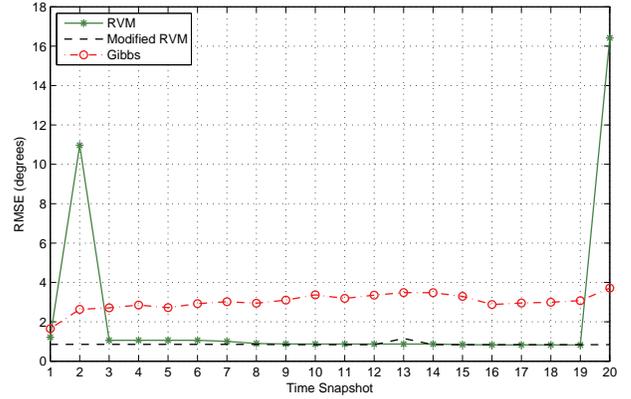}
   \caption{$RMSE$ values for the endfire region example with reduced antenna separation.
    \label{fig:end2}}
\end{center}
\end{figure}

The performances of each of the methods in this instance are summarised in Table \ref{tb:end2} and Figure \ref{fig:end2}, respectively.  Here it can be seen that the larger number of antennas used has resulted in a lower average $RMSE$ values for all three of the methods.  In this instance only the modified RVM method has performed better that the traditional RVM based method when comparing average $RMSE$ values (decrease in average $RMSE$ of 61.09$\%$).  However, by looking at the maximum $RMSE$ values it can be seen that the largest estimation error possible with the traditional RVM based method is larger than that for the Gibbs sampling based method ($16.42^{\circ}$ as compared to $3.71^{\circ}$).

It is worth noting that such an array configuration is unlikely to be used in practice.  This is due to the costs associated with the number of  antennas required.  As a result, the remaining examples will stick to the adjacent antenna separation of $\lambda/2$ and associated parameters previously defined.

\subsection{Non-Endfire Region}
\begin{figure}
\begin{center}
   \includegraphics[angle=0,width=0.45\textwidth]{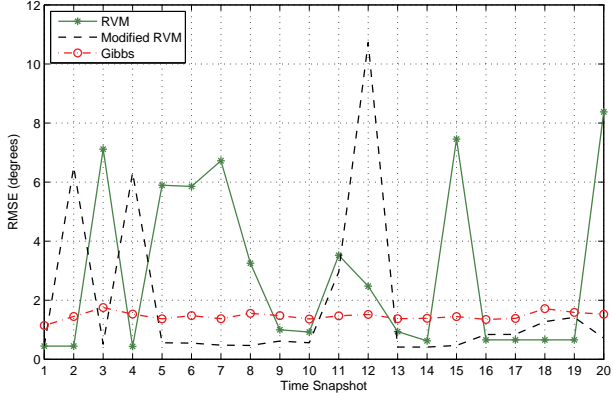}
   \caption{$RMSE$ values for the non-endfire region example.
    \label{fig:non}}
\end{center}
\end{figure}

For this example the initial DOA is $\theta=100^{\circ}$ with the DOA increasing by $1^{\circ}$ at each time snapshot, with the signal value remaining constant at -1.  The performance of the three methods is summarised in Table \ref{tb:non}, with the $RMSE$ values illustrated in Figure \ref{fig:non}.
\begin{table}
\caption{Performance summary for the non-endfire region example.} \centering
\begin{tabular}{|c|c|c|}
  \hline
   & Average $RMSE$  & Average Computation  \\
  Method&(degrees) &Time (seconds) \\
  \hline
  RVM & 2.91 &  0.73\\
  \hline
  Modified RVM &1.85  &  0.78\\
  \hline
  Gibbs & 1.46 &13.08  \text{excluding burn-in}\\
  \cline{3-3}
  &&79.57 \text{including burn-in}\\
  \hline
\end{tabular}
\label{tb:non}
\end{table}

In this instance it can be seen that there has not been as large an increase in $RMSE$ for the traditional RVM method, as the DOA does not enter the endfire region.  However, both the modified RVM and Gibbs sampling based methods have managed to achieve improvements in average $RMSE$ values of 36.42$\%$ and $50.00\%$, respectively.  For the Gibbs sampling based method this comes at the expenses of an increase in computation time, whereas the time for the modified RVM based method is comparable to the traditional RVM based method.    As with the previous test scenario this would suggest that the modified RVM based method should be used when the expected DOA change information is available and the Gibbs sampling based method when this is not the case, or when computational complexity is not a major concern.

\subsection{Random Initial DOA}
Next consider the case where the initial DOA is randomly chosen from the entire angular range and increased by $1^{\circ}$ at each time snapshot.  The signal value is randomly selected as $\pm1$ for each simulation and remains constant as the DOA changes.
\begin{table}
\caption{Performance summary for the random initial DOA example.} \centering
\begin{tabular}{|c|c|c|}
  \hline
   & Average $RMSE$  & Average Computation  \\
  Method&(degrees) &Time (seconds) \\
  \hline
  RVM & 12.10 & 0.74 \\
  \hline
  Modified RVM & 3.22 & 0.86 \\
  \hline
  Gibbs & 2.89 & 13.88 \text{excluding burn-in}\\
  \cline{3-3}
  &&84.19 \text{including burn-in} \\
  \hline
\end{tabular}
\label{tb:ran}
\end{table}

\begin{figure}
\begin{center}
   \includegraphics[angle=0,width=0.45\textwidth]{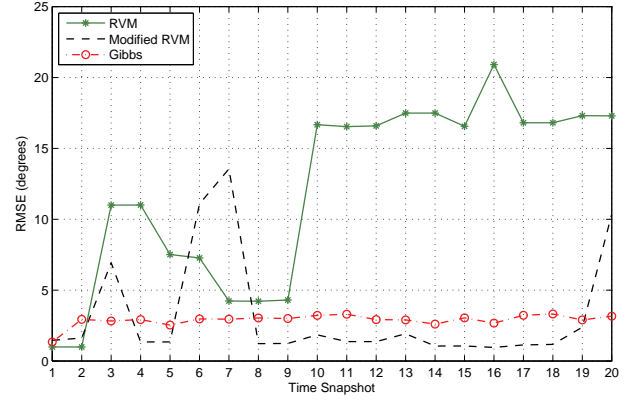}
   \caption{$RMSE$ values for the random initial DOA example.
    \label{fig:random}}
\end{center}
\end{figure}

Table \ref{tb:ran} and Figure \ref{fig:random} summarise the performance of the various methods in this instance.  Again, it can be seen that the modified RVM based method has offered improvements in terms of $RMSE$(73.39$\%$), without a significant increase in computation time.  The Gibbs sampling based method has also given an estimation accuracy improvement (76.12$\%$) and has even outperformed the modified RVM based method, without prior knowledge of how the DOA was going to change.  However, this has come at the expense of an increased computation time.

\subsection{Mismatched Actual and Assumed DOA Change}
This subsection compares the performances of the estimation methods for two situations where the actual change in DOA is not known.  First, consider the case where there is an initial DOA of $\theta=100^{\circ}$ which increases by $1^{\circ}$ for 9 time snapshots before decreasing by $1^{\circ}$ for the remaining time snapshots.  In this instance, assume a constant signal value of 1 throughout.

\begin{figure}
\begin{center}
   \includegraphics[angle=0,width=0.45\textwidth]{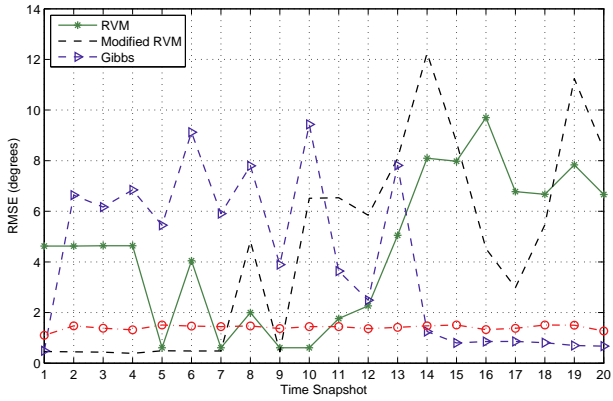}
   \caption{$RMSE$ values for the increasing DOA followed by decreasing DOA example.
    \label{fig:incdec}}
\end{center}
\end{figure}

The performance comparison is now made between the traditional RVM based method, the modified RVM based method with the assumed DOA change set to a constant increase of $1^{\circ}$, the modified RVM based method with the assumed DOA change set to a constant decrease of $1^{\circ}$ and the Gibbs sampling based method.  The performances for each are summarised in Figure \ref{fig:incdec} and Table \ref{tb:incdec}, respectively.

\begin{table}
\caption{Performance summary for the increasing DOA followed by decreasing DOA example.} \centering
\begin{tabular}{|c|c|c|}
  \hline
   & Average $RMSE$  & Average Computation  \\
  Method&(degrees) &Time (seconds) \\
  \hline
  RVM & 4.49 & 0.69 \\
  \hline
  Modified RVM with & 4.45 & 0.77 \\
  constant $+1^{\circ}$ DOA change & &\\
  \hline
  Modified RVM with & 4.08 & 0.79 \\
  constant $-1^{\circ}$ DOA change & & \\
  \hline
  Gibbs & 1.41 & 16.12 \text{excluding burn-in}\\
  \cline{3-3}
  && 97.59 \text{including burn-in} \\
  \hline
\end{tabular}
\label{tb:incdec}
\end{table}

In this instance the average $RMSE$ values suggest a comparable performance in terms of estimation accuracy between the traditional RVM based methods and the two modified RVM based examples.  This can be explained by the fact that for both of the modified RVM based examples, the assumed DOA change does not match the actual DOA changes for the entire time range which means the same improvements as for the previous scenario can no longer be guaranteed.  Figure \ref{fig:incdec} highlights this in the results for the two modified RVM examples.  It demonstrates that with an assumed increasing DOA the modified RVM offers some initial improvements, while the performance is significantly degraded  when the DOA starts to decrease again.  On the other hand, the example with an assumed decreasing DOA performs worse than the traditional RVM based method initially and then offers significant improvements when the actual DOA also starts to decrease.

It can also be seen that for the Gibbs sampling based method there has been an improvement in DOA estimation accuracy.  In terms of average $RMSE$ values this is a decrease of 68.60$\%$, which has been achieved without any knowledge of how the DOA was going to change.  However, there is again an increase in the computation time.

\begin{figure}
\begin{center}
   \includegraphics[angle=0,width=0.45\textwidth]{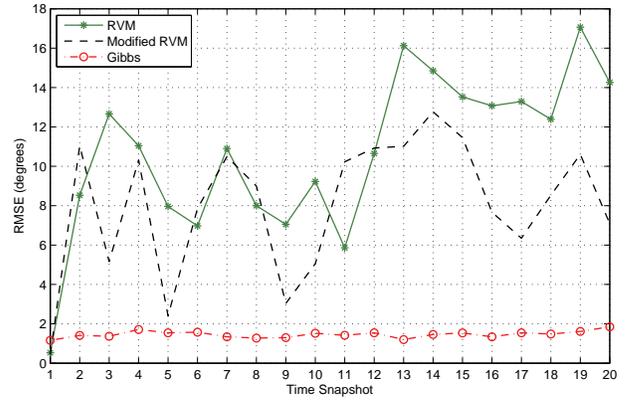}
   \caption{$RMSE$ values for the increasing DOA with an assumed decrease in DOA of $3^{\circ}$ example.
    \label{fig:mismatch}}
\end{center}
\end{figure}

To illustrate how a larger mismatch between actual and assumed DOA changes effects the performance of the modified RVM based method now consider an example where the actual DOA is increasing by $1^{\circ}$ in each snapshot, while the assumed change is a decrease of $3^{\circ}$.  Here, the initial DOA is $100^{\circ}$, with a constant signal value of -1.  The $RMSE$ values for the methods are shown in Figure \ref{fig:mismatch} and summarised in Table \ref{tb:mismatch} along with the computation times.

\begin{table}
\caption{Performance summary for the increasing DOA with an assumed decrease in DOA of $3^{\circ}$ example.} \centering
\begin{tabular}{|c|c|c|}
  \hline
   & Average $RMSE$  & Average Computation  \\
  Method&(degrees) &Time (seconds) \\
  \hline
  RVM & 10.70 & 0.66 \\
  \hline
  Modified RVM& 8.07 & 0.77 \\
  \hline
  Gibbs & 1.46 & 13.99 \text{excluding burn-in}\\
  \cline{3-3}
  && 84.72 \text{including burn-in}\\
  \hline
\end{tabular}
\label{tb:mismatch}
\end{table}

Here it can be seen that the Gibbs sampling based method has offered an $86.36\%$ improvement in average $RMSE$ as compared to the traditional RVM based method.  There has again been a significant increase in the computational complexity.

For the modified RVM based method the average $RMSE$ values suggests that there has been an improvement in estimation accuracy.  However, this is smaller than when the actual and assumed DOA changes match.  It is also unlikely that this improvement would be obtained in every scenario the the method could be applied to.  From looking at Figure \ref{fig:mismatch} we can see that the traditional and modified RVM based methods are showing comparable performance for the just over half of the time frame considered.  This is the relative performance that would be expected in the majority of cases.

\subsection{Random Changes in Direction of Arrival}
\begin{figure}
\begin{center}
   \includegraphics[angle=0,width=0.45\textwidth]{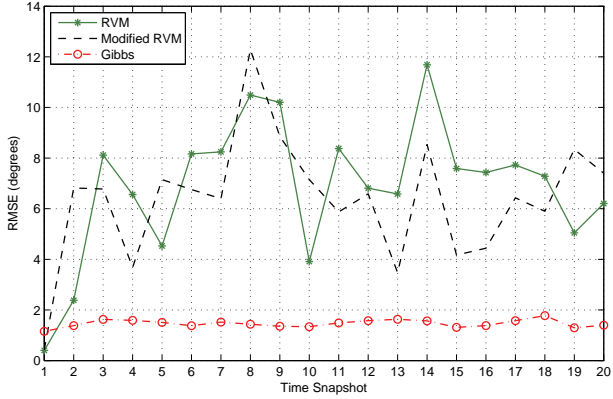}
   \caption{$RMSE$ values for the random DOA change example.
    \label{fig:final}}
\end{center}
\end{figure}

Finally, consider  the example where the initial signal value is assumed to be equal to 1 and the initial DOA is chosen to be $100^{\circ}$.  The actual DOA is then allowed to randomly change by up to $\pm3^{\circ}$ for each time snapshot.  For the modified RVM method  assume that the actual DOA change is an increase of $3^{\circ}$.  This gives the results as summarised in Table \ref{tb:final} and Figure \ref{fig:final}.

It can be seen that the Gibbs sampling based method has again outperformed the modified RVM based method in terms of estimation accuracy, due to the fact that no prior knowledge of how the DOA will change is required.  As compared to the traditional RVM based method there has been an improvement in $RMSE$ of 78.81$\%$.  However, as is expected this is at the cost of computation time.

It is also worth noting that the average $RMSE$ values suggest that the modified RVM and traditional RVM have offered a comparable performance.  This is due to the fact that the assumption of how the DOA will change is not valid, meaning the modified RVM no longer offers any improvements.  Therefore, in this situation the Gibbs sampling based method would be the best to use, assuming computational complexity is not the main motivating factor.
\begin{table}
\caption{Performance summary for the random changes in DOA with an assumed increase in DOA of $3^{\circ}$ example.} \centering
\begin{tabular}{|c|c|c|}
  \hline
   & Average $RMSE$  & Average Computation  \\
  Method&(degrees) &Time (seconds) \\
  \hline
  RVM &6.89  & 0.64 \\
  \hline
  Modified RVM&6.37  & 0.86 \\
  \hline
  Gibbs & 1.46 & 22.26 \text{excluding burn-in}\\
  \cline{3-3}
  && 134.56\text{including burn-in}\\
  \hline
\end{tabular}
\label{tb:final}
\end{table}

\section{Conclusions}\label{sec:con}
This paper has proposed two novel approaches for the
estimation of a dynamic direction of arrival using uniform linear
arrays with mutual coupling. The first approach is a Bayesian
compressed Kalman filter with a modified relevance vector machine,
where the traditional sparsity assumption is replaced by an
assumption that the estimated signals will instead match predicted
signal values. This results in the derivation of a new posterior
probability density function of the received signals and the
expression for the related marginal likelihood function.  The second proposed approach is a Gibbs sampling approach, where sparsity is explicitly enforced if there is a large difference between the previous direction of arrival estimate and the angle currently being considered.  The proposed
approaches will be particularly useful when applied to the problem
of dynamic direction of arrival estimation in the endfire region of antenna arrays.
Such problems can arise in numerous application areas such as in
communications and surveillance.

 An extensive performance evaluation is provided and shows that both of the proposed approaches outperform the traditional relevance vector machine based Bayesian compressive sensing Kalman filter in terms of mean root mean square error values, by up to 73.39$\%$ for the modified relevance vector machine based method and 86.36$\%$ for the Gibbs sampling based method.

\section*{Appendix}
\subsection{Derivation of Posterior Distribution}
Bayes' rule gives
\begin{eqnarray}\label{eq:Bayes}\nonumber
  \mathcal{P}(\tilde{\textbf{x}}_{k}|\tilde{\textbf{y}}_{k},\sigma^{2}_{k},\textbf{p}_{k},\tilde{\textbf{x}}_{p})\mathcal{P}(\tilde{\textbf{y}}_{k}|\sigma^{2}_{k},\textbf{p}_{k},\tilde{\textbf{x}}_{p})&=&\\
  \mathcal{P}(\tilde{\textbf{y}}_{k}|\tilde{\textbf{x}}_{k},\sigma_{k}^{2})\mathcal{P}(\tilde{\textbf{x}}_{k}|\textbf{p}_{k},\tilde{\textbf{x}}_{p}),
\end{eqnarray}
where $\mathcal{P}(\tilde{\textbf{y}}_{k}|\tilde{\textbf{x}}_{k},\sigma_{k}^{2})$ and $\mathcal{P}(\tilde{\textbf{x}}_{k}|\textbf{p}_{k},\tilde{\textbf{x}}_{p})$ are known from (\ref{eq:y dist}) and (\ref{eq:x dist}), respectively.

Now following the method suggested in \cite{Tipping01} carry out the multiplication on the right hand side of \eqref{eq:Bayes}, collect terms in $\tilde{\textbf{x}}_{k}$ in the exponential and complete the square.
\begin{eqnarray}\nonumber
  && -\frac{1}{2}\Big[\sigma_{k}^{-2}(\tilde{\textbf{y}}_{k} - \tilde{\textbf{A}}\tilde{\textbf{x}}_{k})^{T}(\tilde{\textbf{y}}_{k} - \tilde{\textbf{A}}\tilde{\textbf{x}}_{k})+\\ \nonumber &&\;\;\;\;\;\;\;\;\;\;\;\;\;\;\;\;\;\;(\tilde{\textbf{x}}_{k} - \tilde{\textbf{x}}_{p})^{T}\textbf{P}_{k}(\tilde{\textbf{x}}_{k} - \tilde{\textbf{x}}_{p})\Big] \\ \nonumber
  &=& -\frac{1}{2}\Big[ \sigma^{-2}\tilde{\textbf{y}}_{k}^{T}\tilde{\textbf{y}}_{k} - \sigma_{k}^{-2}\tilde{\textbf{y}}_{k}^{T}\tilde{\textbf{A}}\tilde{\textbf{x}}_{k} - \sigma^{-2}\tilde{\textbf{x}}_{k}^{T}\tilde{\textbf{A}}^{T}\tilde{\textbf{y}}_{k} +\\ \nonumber && \sigma_{k}^{-2}\tilde{\textbf{x}}_{k}^{T}\tilde{\textbf{A}}^{T}\tilde{\textbf{A}}\tilde{\textbf{x}}_{k} + \tilde{\textbf{x}}_{k}^{T}\textbf{P}_{k}\tilde{\textbf{x}}_{k} - \tilde{\textbf{x}}_{k}^{T}\textbf{P}_{k}\tilde{\textbf{x}}_{p} - \tilde{\textbf{x}}_{p}^{T}\textbf{P}_{k}\tilde{\textbf{x}}_{k}\\ \nonumber && \;\;\;\;\;\;\;\;\;\;\;\;\;\;\;\;\;\;\;\;\;\;\;\;\;\;\;\;\;\;\;\;\;\;\;\; + \tilde{\textbf{x}}_{p}^{T}\textbf{P}_{k}\tilde{\textbf{x}}_{p} \Big] \\ \nonumber
   &=&  -\frac{1}{2}\Big[ (\tilde{\textbf{x}}_{k}-\boldsymbol\mu)^{T}\boldsymbol\Sigma^{-1}(\tilde{\textbf{x}}_{k}-\boldsymbol\mu) - \boldsymbol\mu^{T}\boldsymbol\Sigma^{-1}\boldsymbol\mu + \\  && \;\;\;\;\;\;\;\;\;\;\;\;\;\;\;\;\;\;\;\;\; \sigma^{-2}\tilde{\textbf{y}}_{k}^{T}\tilde{\textbf{y}}_{k} + \tilde{\textbf{x}}_{p}^{T}\textbf{P}_{k}\tilde{\textbf{x}}_{p} \Big]
\end{eqnarray}
where $\boldsymbol\Sigma$ and $\boldsymbol\mu$ are given by (\ref{eq:SIGMA}) and (\ref{eq:mu}), respectively. This then gives the posterior distribution as (\ref{eq:post}), with the remaining exponential terms
\begin{equation}\label{eq:remain1}
  -\frac{1}{2}\Bigg[ \sigma_{k}^{-2}\tilde{\textbf{y}}_{k}^{T}\tilde{\textbf{y}}_{k} + \tilde{\textbf{x}}_{p}^{T}\textbf{P}_{k}\tilde{\textbf{x}}_{p} - \boldsymbol\mu^{T}\boldsymbol\Sigma^{-1}\boldsymbol\mu  \Bigg].
\end{equation}
\subsection{Derivation of Marginal Likelihood}
From (\ref{eq:Bayes}) the following is known:
\begin{equation}\label{eq:Bayes2}
  \mathcal{P}(\tilde{\textbf{y}}_{k}|\sigma^{2}_{k},\textbf{p}_{k},\tilde{\textbf{x}}_{p})=\frac{\mathcal{P}(\tilde{\textbf{y}}_{k}|\tilde{\textbf{x}}_{k},\sigma_{k}^{2}),\mathcal{P}(\tilde{\textbf{x}}_{k}|\textbf{p}_{k},\tilde{\textbf{x}}_{p})}{\mathcal{P}(\tilde{\textbf{x}}_{k}|\tilde{\textbf{y}}_{k},\sigma^{2}_{k},\textbf{p}_{k},\tilde{\textbf{x}}_{p})},
\end{equation}
meaning the term in the exponential will be (\ref{eq:remain1}) where
\begin{eqnarray}\label{eq:musigmu}\nonumber
  \boldsymbol\mu^{T}\boldsymbol\Sigma^{-1}\boldsymbol\mu&=&(\sigma_{k}^{-2}\tilde{\textbf{A}}^{T}\tilde{\textbf{y}}_{k}+\textbf{P}_{k}\tilde{\textbf{x}}_{p})^{T}\boldsymbol\Sigma^{T}\boldsymbol\Sigma^{-1}\\ \nonumber &\times&\boldsymbol\Sigma(\sigma_{k}^{-2}\tilde{\textbf{A}}^{T}\tilde{\textbf{y}}_{k}+\textbf{P}_{k}\tilde{\textbf{x}}_{p})\\ \nonumber
  &=&(\sigma_{k}^{-2}\tilde{\textbf{A}}^{T}\tilde{\textbf{y}}_{k}+\textbf{P}_{k}\tilde{\textbf{x}}_{p})^{T}
  (\sigma_{k}^{-2}\boldsymbol\Sigma\tilde{\textbf{A}}^{T}\tilde{\textbf{y}}_{k}+\boldsymbol\Sigma\textbf{P}_{k}\tilde{\textbf{x}}_{p})\\ \nonumber &=&\sigma_{k}^{-4}\tilde{\textbf{y}}_{k}^{T}\tilde{\textbf{A}}\boldsymbol\Sigma\tilde{\textbf{A}}^{T}\tilde{\textbf{y}}_{k} + \sigma^{-2}\tilde{\textbf{y}}_{k}^{T}\tilde{\textbf{A}}\boldsymbol\Sigma\textbf{P}_{k}\tilde{\textbf{x}}_{p} + \\ && \sigma_{k}^{-2}\tilde{\textbf{x}}_{p}^{T}\textbf{P}_{k}^{T}\tilde{\textbf{A}}^{T}\tilde{\textbf{y}}_{k} + \tilde{\textbf{x}}_{p}^{T}\textbf{P}_{k}^{T}\boldsymbol\Sigma\textbf{P}_{k}\tilde{\textbf{x}}_{p}.
\end{eqnarray}
Therefore the exponential term is given by
\begin{eqnarray}\label{eq:exp}\nonumber
  -\frac{1}{2}\Bigg[\sigma_{k}^{-2}\tilde{\textbf{y}}_{k}^{T}\tilde{\textbf{y}}_{k} + \tilde{\textbf{x}}_{p}^{T}\textbf{P}_{k}\tilde{\textbf{x}}_{p} - \sigma_{k}^{-4}\tilde{\textbf{y}}_{k}^{T}\tilde{\textbf{A}}\boldsymbol\Sigma\tilde{\textbf{A}}^{T}\tilde{\textbf{y}}_{k} - \\ \nonumber  \sigma_{k}^{-2}\tilde{\textbf{y}}_{k}^{T}\tilde{\textbf{A}}\boldsymbol\Sigma\textbf{P}_{k}\tilde{\textbf{x}}_{p} - \sigma_{k}^{-2}\tilde{\textbf{x}}_{p}^{T}\textbf{P}_{k}^{T}\tilde{\textbf{A}}^{T}\tilde{\textbf{y}}_{k} -  \tilde{\textbf{x}}_{p}^{T}\textbf{P}_{k}^{T}\boldsymbol\Sigma\textbf{P}_{k}\tilde{\textbf{x}}_{p}   \Bigg] \\ \nonumber
  =  -\frac{1}{2}\Bigg[\tilde{\textbf{y}}_{k}^{T}[\sigma_{k}^{-2}-\sigma_{k}^{-4}\tilde{\textbf{A}}\boldsymbol\Sigma\tilde{\textbf{A}}^{T}]
   \tilde{\textbf{y}}_{k} + \tilde{\textbf{x}}_{p}^{T}[\textbf{P}_{k}-\textbf{P}_{k}^{T}\boldsymbol\Sigma\textbf{P}_{k}]\tilde{\textbf{x}}_{p}\\   -   \sigma_{k}^{-2}\tilde{\textbf{y}}_{k}^{T}\tilde{\textbf{A}}\boldsymbol\Sigma\textbf{P}_{k}\tilde{\textbf{x}}_{p} - \sigma_{k}^{-2}\tilde{\textbf{x}}_{p}^{T}\textbf{P}_{k}^{T}\tilde{\textbf{A}}^{T}\tilde{\textbf{y}}_{k}  \Bigg]
\end{eqnarray}

The term outside of the exponential is given by
\begin{equation}\label{eq:nonexp}
  \frac{(2\pi\sigma_{k}^{2})^{-M}(2\pi)^{-N}|\textbf{P}_{k}|^{1/2}}{(2\pi)^{-N}|\boldsymbol\Sigma|^{-\frac{1}{2}}}
=(2\pi\sigma_{k}^{2})^{-M}|\boldsymbol\Sigma|^{\frac{1}{2}}|\textbf{P}_{k}|^{\frac{1}{2}}.
\end{equation}
This gives the marginal likelihood as
\begin{eqnarray}\label{eq:marg}\nonumber
  \mathcal{P}(\tilde{\textbf{y}}_{k}|\sigma^{2}_{k},\textbf{p}_{k},\tilde{\textbf{x}}_{p})=(2\pi\sigma_{k}^{2})^{-M}|\boldsymbol\Sigma|^{\frac{1}{2}}|\textbf{P}_{k}|^{\frac{1}{2}}\\ \times\exp\Big\{ -\frac{1}{2}[\tilde{\textbf{y}}_{k}^{T}\textbf{B}\tilde{\textbf{y}}_{k}+\tilde{\textbf{x}}_{p}^{T}\textbf{C}\tilde{\textbf{x}}_{p}-\;2\sigma_{k}^{2}\tilde{\textbf{y}}_{k}^{T}\tilde{\textbf{A}}\boldsymbol\Sigma\textbf{P}_{k}\tilde{\textbf{x}}_{p}]\Big\},
\end{eqnarray}
where $\textbf{B}$ and $\textbf{C}$ are defined as in Section \ref{sub:RVM}.  The log likelihood is then given by
\begin{eqnarray}\label{eq:log2}\nonumber
  \mathcal{L}(\sigma^{2}_{k},\textbf{p}_{k}) = \log\Bigg\{ (2\pi\sigma_{k}^{2})^{-M}|\boldsymbol\Sigma|^{\frac{1}{2}}|\textbf{P}_{k}|^{\frac{1}{2}}\\ \nonumber\times\exp\Big\{ -\frac{1}{2}[\tilde{\textbf{y}}_{k}^{T}\textbf{B}\tilde{\textbf{y}}_{k}+\tilde{\textbf{x}}_{p}^{T}\textbf{C}\tilde{\textbf{x}}_{p}-2\sigma_{k}^{2}\tilde{\textbf{y}}_{k}^{T}\tilde{\textbf{A}}\boldsymbol\Sigma\textbf{P}_{k}\tilde{\textbf{x}}_{p}]\Big\} \Bigg\} \\  \nonumber
  = -M\log(2\pi) -M\log\sigma_{k}^{2} +\frac{1}{2}\log|\boldsymbol\Sigma| +\\  \frac{1}{2}\log|\textbf{P}_{k}| -\frac{1}{2}[\tilde{\textbf{y}}_{k}^{T}\textbf{B}\tilde{\textbf{y}}_{k}+\tilde{\textbf{x}}_{p}^{T}\textbf{C}\tilde{\textbf{x}}_{p}-2\sigma_{k}^{2}\tilde{\textbf{y}}_{k}^{T}\tilde{\textbf{A}}\boldsymbol\Sigma\textbf{P}_{k}\tilde{\textbf{x}}_{p}].
\end{eqnarray}

Using the Woodbury matrix inversion identity gives
\begin{equation}\label{eq:invB}
  \textbf{B} = \sigma_{k}^{-2}\textbf{I}-\sigma_{k}^{-2}\tilde{\textbf{A}}(\textbf{P}_{k}+\sigma_{k}^{-2}\tilde{\textbf{A}}^{T}\tilde{\textbf{A}})^{-1}\tilde{\textbf{A}}^{T}\sigma_{k}^{-2},
\end{equation}
which means
\begin{eqnarray}\nonumber
  \tilde{\textbf{y}}_{k}^{T}\textbf{B}\tilde{\textbf{y}}_{k} &=& \tilde{\textbf{y}}_{k}^{T}\sigma_{k}^{-2}\tilde{\textbf{y}}_{k} - \tilde{\textbf{y}}_{k}^{T}(\sigma_{k}^{-2}\textbf{I}-\sigma_{k}^{-2}\tilde{\textbf{A}}\\ \nonumber &\times&(\textbf{P}+\sigma_{k}^{-2}\tilde{\textbf{A}}^{T}\tilde{\textbf{A}})^{-1}\tilde{\textbf{A}}^{T}\sigma_{k}^{-2})\tilde{\textbf{y}}_{k} \\ \nonumber
   &=&  \tilde{\textbf{y}}_{k}^{T}\sigma_{k}^{-2}\tilde{\textbf{y}}_{k} - \tilde{\textbf{y}}_{k}^{T}\sigma_{k}^{-2}\tilde{\textbf{A}}\boldsymbol\Sigma\tilde{\textbf{A}}^{T}\sigma_{k}^{-2}\tilde{\textbf{y}}_{k} \\ \nonumber
   &=& \sigma_{k}^{-2}\tilde{\textbf{y}}_{k}^{T}(\tilde{\textbf{y}}_{k}-\tilde{\textbf{A}}\boldsymbol\mu)+\sigma_{k}^{-2}\tilde{\textbf{y}}_{k}^{T}\tilde{\textbf{A}}\boldsymbol\Sigma\textbf{P}_{k}\tilde{\textbf{x}}_{p} \\ \nonumber
   &=&  \sigma_{k}^{-2}||\tilde{\textbf{y}}_{k}^{T}-\tilde{\textbf{A}}\boldsymbol\mu||_{2}^{2}+\boldsymbol\mu^{T}\textbf{P}_{k}\boldsymbol\mu+\sigma^{-2}\tilde{\textbf{y}}_{k}^{T}\tilde{\textbf{A}}\boldsymbol\Sigma\textbf{P}_{k}\tilde{\textbf{x}}_{p}.\\
\end{eqnarray}
Also, we know that $\textbf{P}_{k}^{T}=\textbf{P}_{k}$ as $\textbf{P}_{k}$ is a real valued diagonal matrix.  This means
\begin{eqnarray}\nonumber
  \tilde{\textbf{x}}_{p}^{T}\textbf{C}\tilde{\textbf{x}}_{p} &=& \tilde{\textbf{x}}_{p}^{T}[\textbf{P}_{k}-\textbf{P}_{k}\boldsymbol\Sigma\textbf{P}_{k}]\tilde{\textbf{x}}_{p} \\ \nonumber
   &=& \tilde{\textbf{x}}_{p}^{T}\textbf{P}_{k}\tilde{\textbf{x}}_{p} -\tilde{\textbf{x}}_{p}^{T}\textbf{P}_{k}\boldsymbol\Sigma\textbf{P}_{k}\tilde{\textbf{x}}_{p}\\
   &=&  \tilde{\textbf{x}}_{p}^{T}\textbf{P}_{k}\tilde{\textbf{x}}_{p} - \tilde{\textbf{x}}_{p}^{T}\textbf{P}_{k}\boldsymbol\mu+\sigma_{k}^{-2}\tilde{\textbf{y}}_{k}^{T}\tilde{\textbf{A}}\boldsymbol\Sigma\textbf{P}_{k}\tilde{\textbf{x}}_{p},
\end{eqnarray}
which then gives the log likelihood function in (\ref{eq:log}).
\subsection{Derivation of Update Expressions for Modified RVM}
Firstly, differentiating (\ref{eq:log}) with respect to $p_{k,n}$ gives
\begin{equation}\label{eq:p1}
  -\frac{1}{2}\Big[\Sigma_{nn}-\frac{1}{p_{k,n}}+\mu^{2}_{n}+\tilde{x}_{e,n}^{2}-\tilde{x}_{e,n}\mu_{n} \Big]
\end{equation}
and equating to zero gives
\begin{eqnarray}\label{eq:p2}\nonumber
  \Sigma_{nn}-\frac{1}{p_{k,n}}+\mu^{2}_{n}+\tilde{x}_{e,n}^{2}-\tilde{x}_{e,n}\mu_{n} = 0 &&\\ \nonumber
  1 - p_{k,n}\Sigma_{nn}-p_{k,n}\mu_{n}^{2}-p_{k,n}\tilde{x}_{e,n}^{2}+p_{k,n}\tilde{x}_{e,n}\mu_{n} = 0 &&\\
  \gamma_{n}-p_{k,n}[\mu_{n}^{2}+\tilde{x}_{e,n}^{2}-\tilde{x}_{e,n}\mu_{n}]=0&&
\end{eqnarray}
which leads to (\ref{eq:pnew}).

Now collect the terms with $\sigma_{k}^{2}$ in to give
\begin{equation}\label{eq:sig1}
  -\frac{1}{2}\Big[ 2M\log\sigma_{k}^{2} - \log|\boldsymbol\Sigma| + \sigma_{k}^{-2}||\tilde{\textbf{y}}_{k}-\tilde{\textbf{A}}\boldsymbol\mu||_{2}^{2} \Big]
\end{equation}
and then define $\tau=\sigma_{k}^{-2}$ giving
\begin{eqnarray}\label{eq:sig2}\nonumber
  && -\frac{1}{2}\Big[ 2M\log\tau^{-1} - \log|\boldsymbol\Sigma| + \tau||\tilde{\textbf{y}}_{k}-\tilde{\textbf{A}}\boldsymbol\mu||_{2}^{2} \Big] \\
  &=& -\frac{1}{2}\Big[ -2M\log\tau - \log|\boldsymbol\Sigma| + \tau||\tilde{\textbf{y}}_{k}-\tilde{\textbf{A}}\boldsymbol\mu||_{2}^{2} \Big].
\end{eqnarray}
Now differentiate (\ref{eq:sig2}) with respect to $\tau$ and equate to zero to give
\begin{equation}\label{eq:sig3}
  -\frac{2M}{\tau}+\text{tr}(\boldsymbol{\Sigma}\tilde{\textbf{A}}^{T}\tilde{\textbf{A}})+||\tilde{\textbf{y}}_{k}-\tilde{\textbf{A}}\boldsymbol\mu||_{2}^{2} = 0,
\end{equation}
where $\text{tr}(\cdot)$ indicates the trace.  As
$\text{tr}(\boldsymbol{\Sigma}\tilde{\textbf{A}}^{T}\tilde{\textbf{A}})$
can be written as $\tau^{-1}\sum\limits_{n}\gamma_{n}$ giving
\begin{equation}\label{eq:sig4}
  \tau^{-1}(2M-\sum\limits_{n}\gamma_{n})=||\tilde{\textbf{y}}_{k}-\tilde{\textbf{A}}\boldsymbol\mu||_{2}^{2},
\end{equation}
which in turn gives (\ref{eq:sigmanew}).

\subsection{Derivation of \eqref{eq:gibbs1}, \eqref{eq:gibbs2}, \eqref{eq:gibbs3} and \eqref{eq:gibbs4}}
From \eqref{eq:postx} it is known that
\begin{eqnarray}\label{eq:D1}\nonumber
  \mathcal{P}(\tilde{x}_{k,n}|\tilde{y}_{k,n},\sigma_{k}^{2},p_{k,n},\tilde{z}_{k,n})\propto(1-\tilde{z}_{k,n})\delta_{0}\\ \nonumber\times\mathcal{N}(\tilde{\textbf{y}}_{k,n}|\tilde{\textbf{A}}_{n}\tilde{x}_{k,n},\sigma_{k}^{2})\\ +\tilde{z}_{k,n}\mathcal{N}(\tilde{x}_{k,n}|0,p_{k,n})\mathcal{N}(\tilde{\textbf{y}}_{k,n}|\tilde{\textbf{A}}_{n}\tilde{x}_{k,n},\sigma_{k}^{2}).
\end{eqnarray}

If we then combine the exponential terms in the second term in \eqref{eq:D1} we get
\begin{eqnarray}\nonumber
    &&-\frac{1}{2}\Bigg[\tilde{x}_{k,n}^{T}p_{k,n}\tilde{x}_{k,n}+p_{0}(\tilde{\textbf{y}}_{k,n}-\tilde{\textbf{A}}_{n}\tilde{x}_{k,n})^{T}\\ \nonumber&&\;\;\;\;\; \times(\tilde{\textbf{y}}_{k,n}-\tilde{\textbf{A}}_{n}\tilde{x}_{k,n})\Bigg]   \\ \nonumber
    &=& -\frac{1}{2}\Bigg[\tilde{x}_{k,n}^{T}p_{k,n}\tilde{x}_{k,n}+p_{0}(\tilde{\textbf{y}}_{k,n}^{T}\tilde{\textbf{y}}_{k,n}-\tilde{\textbf{y}}_{k,n}^{T}\tilde{\textbf{A}}_{n}\tilde{x}_{k,n}\\ \nonumber&&\;\;\;\;\;\;-\tilde{x}_{k,n}^{T}\tilde{\textbf{A}}_{n}^{T}\tilde{\textbf{y}}_{k,n}+\tilde{x}_{k,n}^{T}\tilde{\textbf{A}}_{n}^{T}\tilde{\textbf{A}}_{n}\tilde{x}_{k,n})\Bigg] \\ \nonumber
    &=&-\frac{1}{2}\Bigg[\tilde{x}_{k,n}^{T}[p_{k,n}+p_{0}\tilde{\textbf{A}}_{n}^{T}\tilde{\textbf{A}}_n]\tilde{x}_{k,n}+p_{0}\tilde{\textbf{y}}_{k,n}^{T}\tilde{\textbf{y}}_{k,n}-\\  &&\;\;\;\;\;p_{0}\tilde{\textbf{y}}_{k,n}^{T}\tilde{\textbf{A}}_{n}\tilde{x}_{k,n}-p_{0}\tilde{x}_{k,n}^{T}\tilde{\textbf{A}}_{n}^{T}\tilde{\textbf{y}}_{k,n}\Bigg].
\end{eqnarray}
Completing the square gives
\begin{eqnarray}\label{eq:D2}\nonumber
    -\frac{1}{2}\Bigg[(\tilde{x}_{k,n}-\hat{\mu}_{n})^{T}\hat{p}_{k,n}(\tilde{x}_{k,n}-\hat{\mu}_{n})-\hat{\mu}_{n}\hat{p}_{k,n}\hat{\mu}_{n} \\  \;\;\;\;\;\;\;\;\;\;\;\;\;\;\;\;\;\;\;\;\;\;\;\;\;\;\;\;\;\;+p_{0}\tilde{\textbf{y}}_{k,n}^{T}\tilde{\textbf{y}}_{k,n}\Bigg],
\end{eqnarray}
where $\hat{p}_{k,n}$ and $\hat{\mu}_{k,n}$ are given by \eqref{eq:gibbs2} and \eqref{eq:gibbs3}, respectively.

In order to complete the expression given in \eqref{eq:gibbs1} it is now
necessary to get a new indicator variable, $\hat{z}_{k,n}$ for the new
posterior distribution for $\tilde{x}_{k,n}$.  To do this, assume
that
\begin{equation}\label{eq:D3}
  \frac{\tilde{z}_{k,n}}{1-\tilde{z}_{k,n}}\mathcal{N}(0|0,p_{k,n}) = \frac{\hat{z}_{k,n}}{1-\hat{z}_{k,n}}\mathcal{N}(0|\hat{\mu}_{n},\hat{p}_{k,n}).
\end{equation}
Thus giving us \eqref{eq:gibbs4}, allowing us to write the posterior distribution for $x_{k,n}$ in the form given in \eqref{eq:gibbs1}.
\section*{Acknowledgments}

\noindent We appreciate the support of the UK Engineering and Physical Sciences Research Council (EPSRC) via the project Bayesian Tracking and Reasoning over Time (BTaRoT) grant EP/K021516/1.  We acknowledge the anonymous reviewers' suggestions that have helped improve this work and would like to thank the associate editor for handling the review of our paper.

\newpage
\begin{IEEEbiography}
[{\includegraphics[width=1.29in,height=1.39in,trim = 15mm 14mm -10mm
0mm]{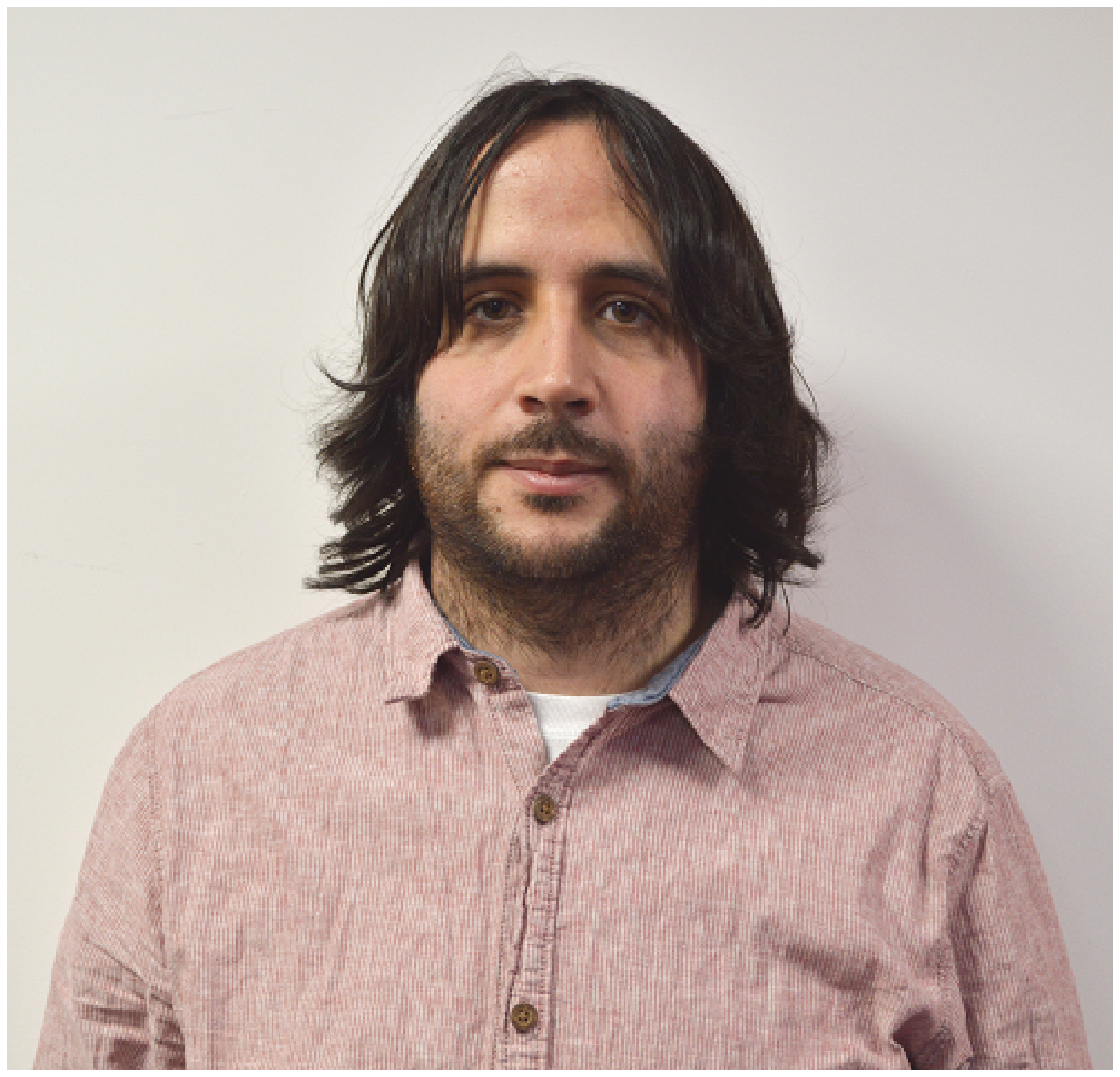}}] {Matthew Hawes} received his MEng and
PhD degree from the Department of Electronic and Electrical
Engineering at the University of Sheffield in 2010 and 2014,
respectively.  Since then he has been employed as a research
associate in the Department of Automatic Control and Systems
Engineering at the same university.  He is currently working on the
EU funded SETA project, the main scope of which is the development of
models, methods and a platform for mobility prediction, congestion
avoidance and sensor data fusion for smart cities. His research
interests include array signal processing, localisation and
tracking, big data, modelling complex systems, intelligent
transportation systems, mobility, data fusion, sequential Monte
Carlo methods and Markov chain Monte Carlo methods.
\end{IEEEbiography}
\vspace*{-1cm}
\begin{IEEEbiography}
[{\includegraphics[width=1.39in,height=1.35in,trim = 15mm 14mm -10mm
0mm]{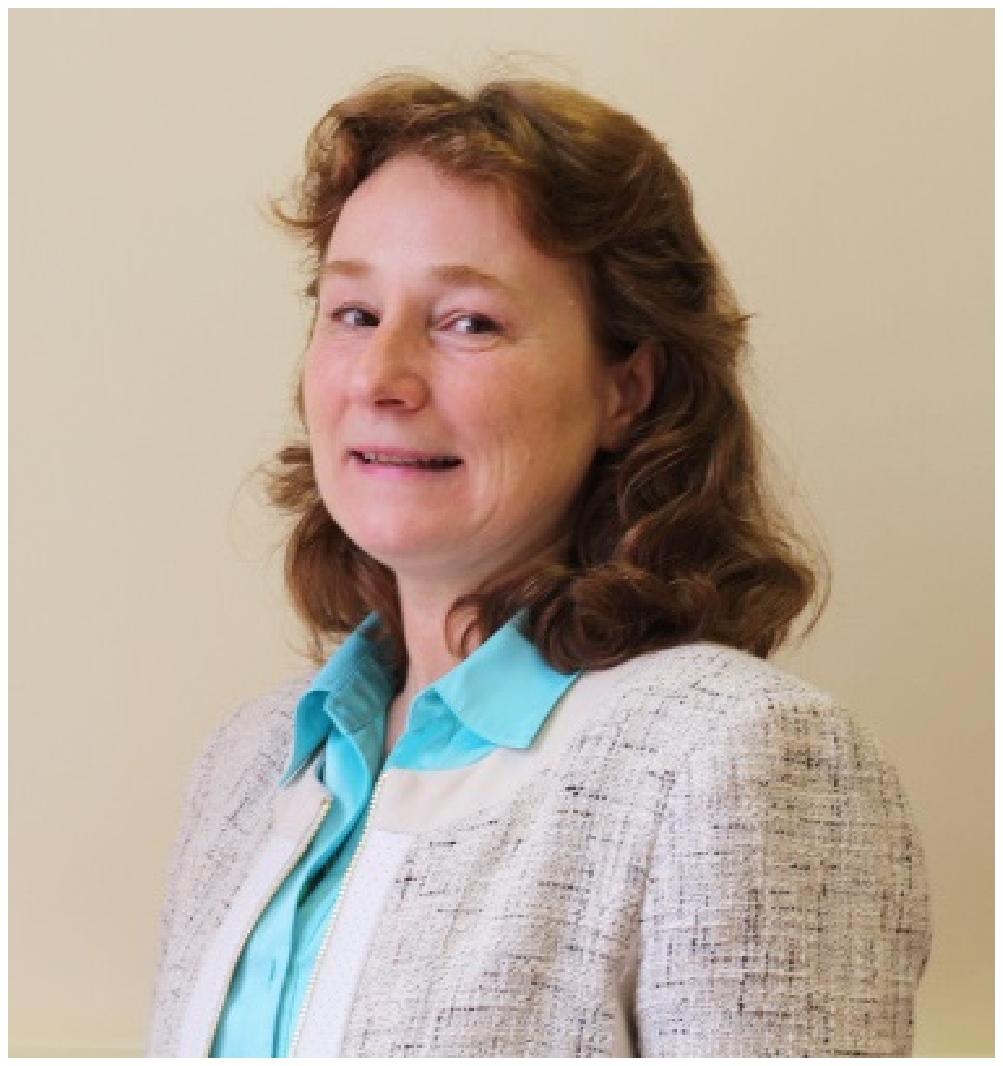}}] {Lyudmila Mihaylova} (M'98,
SM'2008) is Professor of Signal Processing and Control at the
Department of Automatic Control and Systems Engineering at the
University of Sheffield, United Kingdom. Her research is in the
areas of machine learning and autonomous systems with various
applications such as navigation, surveillance and sensor network
systems. She has given a number of talks and tutorials, including
the plenary talk for the IEEE Sensor Data Fusion 2015 (Germany),
invited talks University of California, Los Angeles, IPAMI Traffic
Workshop 2016 (USA), IET ICWMMN 2013 in Beijing, China. Dr.
Mihaylova is an Associate Editor of the IEEE Transactions on
Aerospace and Electronic Systems and of the Elsevier Signal
Processing Journal. She was elected in March 2016 as a president of
the International Society of Information Fusion (ISIF). She is on
the board of Directors of ISIF and a Senior IEEE member. She was the
general co-chair IET Data Fusion $\&$ Target Tracking 2014 and 2012
Conferences, Program co-chair for the 19th International Conference
on Information Fusion, Heidelberg, Germany, 2016, academic chair of
Fusion 2010 conference.
\end{IEEEbiography}
\vspace*{-1cm}
\begin{IEEEbiography}
[{\includegraphics[width=1.0in,height=1.25in,clip]{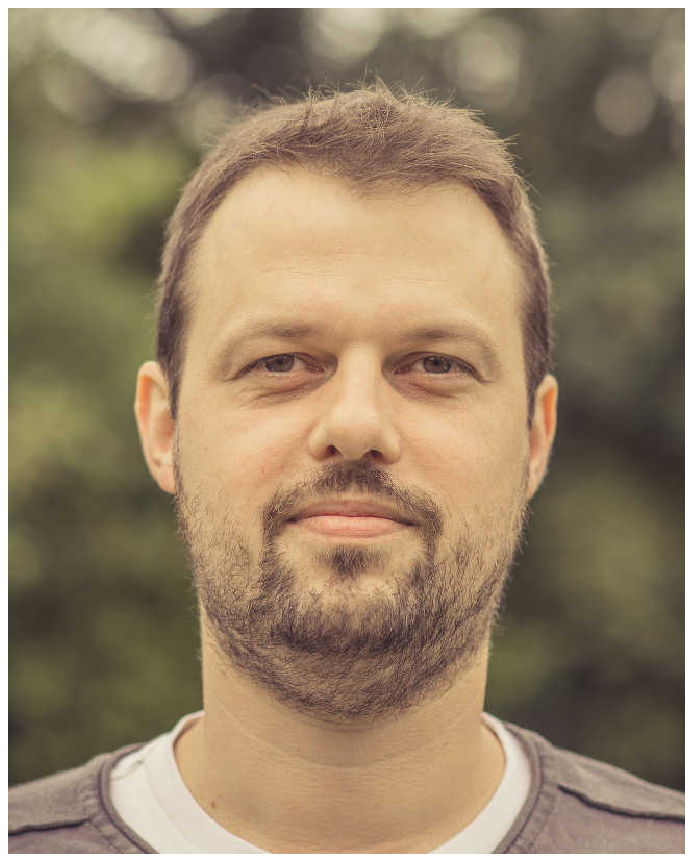}}]
{Fran\c{c}ois Septier} received the Engineer Degree in electrical
engineering and signal processing in 2004 from T\'el\'ecom Lille
France, and a Ph.D in Electrical Engineering from the University of
Valenciennes France in 2008. From March 2008 to August 2009, he was
a Research Associate in the Signal Processing and Communications
Laboratory, Cambridge University, Engineering Department, UK. From
August 2009, he is an Associate Professor with the IMT Lille Douai /
CRIStAL UMR CNRS 9189, France. His research focuses on Bayesian
computational methodology with a particular emphasis on the
development of Monte Carlo based approaches for complex and
high-dimensional problems.
\end{IEEEbiography}
\vspace*{-1cm}
\begin{IEEEbiography}
[{\includegraphics[width=1.0in,height=1.25in,clip]{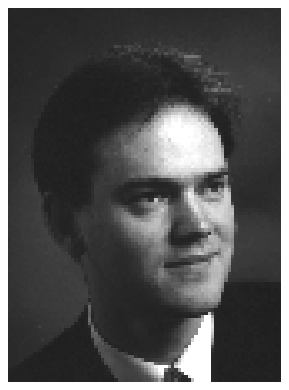}}]
{Simon Godsill} is Professor of Statistical Signal Processing in the
Engineering Department at Cambridge University. He is also a
Professorial Fellow and tutor at Corpus Christi College Cambridge.
He coordinates an active research group in Signal Inference and its
Applications within the Signal Processing and Communications
Laboratory at Cambridge, specializing in Bayesian computational
methodology, multiple object tracking, audio and music processing,
and financial time series modeling. A particular methodological
theme over recent years has been the development of novel techniques
for optimal Bayesian filtering, using Sequential Monte Carlo or
Particle Filtering methods. Prof. Godsill has published extensively
in journals, books and international conference proceedings, and has
given a number of high profile invited and plenary addresses at
conferences such as the Valencia conference on Bayesian Statistics
and the IEEE Statistical Signal Processing Workshop. He was
technical chair of the successful IEEE NSSPW workshop in 2006 on
sequential and nonlinear filtering methods, and has been on the
conference panel for numerous other conferences/workshops. Prof.
Godsill has served as Associate Editor for IEEE Tr. Signal
Processing and the journal Bayesian Analysis. He was Theme Leader in
Tracking and Reasoning over Time for the UK's Data and Information
Fusion Defence Technology Centre (DIF-DTC) and Principal
Investigator on many grants funded by the EU, EPSRC, QinetiQ , MOD,
Microsoft UK, Citibank and Mastercard.
\end{IEEEbiography}
\end{document}